\documentclass{jfm}

\usepackage{graphicx}
\usepackage{tikz}
\usepackage{placeins}
\usepackage{newtxtext}
\usepackage{newtxmath}
\usepackage{natbib}
\usepackage{hyperref}
\hypersetup{
    colorlinks = true,
    urlcolor   = blue,
    citecolor  = black,
}

\newcommand{\RomanNumeralCaps}[1]
\linenumbers

\DeclareRobustCommand\sampleline[1]{%
  \tikz\draw[#1] (0,0) (0,\the\dimexpr\fontdimen22\textfont2\relax)
  -- (2em,\the\dimexpr\fontdimen22\textfont2\relax);%
}

\def\bnab{\mbox{\boldmath $ \nabla$}}
\newcommand{\bu}{\mathbf u}
\newcommand{\ba}{\mathbf a}
\newcommand{\C}{\mathbf C}
\newcommand{\F}{\mathbf F}

\newcommand{\T}{\mathbf T}
\def\bvarphi{\mbox{\boldmath $ \varphi$}}
\def\bgamma{\mbox{\boldmath $ \gamma$}}

\def\b_eta{\mbox{\boldmath $ \eta$}}
\def\tpsi{\psi}

\title{Finite-amplitude elastic waves in viscoelastic channel flow from large to zero Reynolds number}

\author{Gergely Buza\aff{1}
  \corresp{{gb643@cam.ac.uk}},
  Miguel Beneitez\aff{1}\corresp{{mb2467@cam.ac.uk}},
  Jacob Page\aff{2}\corresp{{jacob.page@ed.ac.uk}},
 \and Rich R. Kerswell\aff{1}\corresp{{r.r.kerswell@damtp.cam.ac.uk}}}

\affiliation{
\aff{1}DAMTP, Centre for Mathematical Sciences, Wilberforce Road, Cambridge CB3 0WA, UK
\aff{2}School of Mathematics, University of Edinburgh, EH9 3FD, UK}

\begin{document}
\maketitle

\begin{abstract}

Using branch continuation in the FENE-P model, we show that finite-amplitude travelling waves borne out of the recently-discovered linear instability of viscoelastic channel flow (Khalid et al. {\em J. Fluid Mech.} {\bf 915}, A43, 2021) are substantially subcritical reaching much lower Weissenberg ($Wi$) numbers than on the neutral curve at a given Reynolds ($Re$) number over $Re \in [0,3000]$.  The travelling waves on the lower branch are surprisingly weak indicating that viscolastic channel flow is susceptible to (nonlinear) instability triggered by  small finite amplitude disturbances for $Wi$ and $Re$ well below the neutral curve. The critical $Wi$ for these waves to appear in a saddle node bifurcation decreases monotonically from, for example, $\approx 37$ at $Re=3000$ down to $\approx 7.5$ at $Re=0$ at the solvent-to-total-viscosity ratio $\beta=0.9$. In this latter creeping flow limit, we also show that these waves exist at $Wi \lesssim 50$ for higher polymer concentrations - $\beta \in [0.5,0.97)$ -- where there is no known linear instability. Our results therefore indicate that these travelling waves -- found in simulations and named `arrowheads' by Dubief et al. {\em arXiv}.2006.06770 (2020) - exist much more generally in $(Wi,Re, \beta)$ parameter space than their spawning neutral curve and hence can either directly, or indirectly through their instabilities, influence the dynamics seen far away from where the flow is linearly unstable. Possible connections to elastic and elasto-inertial turbulence are discussed.

\end{abstract}




\section{Introduction}
\label{sect:intro}

It is now well known that even small concentrations of long-chain polymers in a Newtonian solvent can give rise to interesting new behaviour \citep[e.g.][]{Larson1988}. Perhaps the most extreme demonstration of this is the existence of `Elastic' turbulence (ET) at vanishingly small Reynolds numbers ($Re$) where inertia is minimal \citep{Groisman2000,Groisman2001,Steinberg2021}. In 2013, a further multiscale, time-dependent state - `elasto-inertial' turbulence or EIT -  was found which differs from Newtonian turbulence (NT) in being predominantly 2D and seems to require finite Reynolds number ($Re=O(10^3)$) and Weissenberg number $Wi=O(10)$ to exist \citep{Samanta2013, Dubief2013, Sid2018}. Understanding exactly how these different types of turbulence relate to each other remains an outstanding challenge.
Work at the NT-EIT interface  has so far focussed on the possible sustenance of elastically-modified Tollmein-Schlicting waves at least for very dilute solutions and weak elasticity \citep{Shekar2018,Shekar2020}. Our focus here is the possible relationship between EIT and ET: are they two extremes of one whole \citep{Samanta2013, Qin2019, Choueiri2021, Steinberg2021} or distinct flow responses \citep[e.g. see figure 30][]{Chaudhary2021} and figures 21 \& 22 of \cite{Datta2021}?
Finding the dynamical origin for either could help in resolving this question.

The very recent discovery of a new linear instability in dilute viscoelastic rectilinear flows at high $Wi=O(20)$ (in pipes by \cite{Garg2018} and channels by \cite{Khalid2021a}) seems highly relevant. Such `straight' flows had always been believed linearly stable due to the absence of curved streamlines \citep[e.g. see][for extensive discussion of this]{Chaudhary2019, Chaudhary2021,Datta2021, CastilloSanchez2022} although there had been some  evidence of instability to finite-amplitude disturbances at low $Re$ \citep{Bertola2003,Pan2013,Choueiri2021,jha2020}. 
Significantly, the neutral curve for this instability lies in a region of the $(Wi,Re)$ parameter space between where EIT and ET are believed to exist.
The instability was initially only found above $Re \approx 63$ in pipe flow in the Oldroyd-B model \citep{Garg2018, Chaudhary2021}, suggesting that it needs some inertia to function. 
However, the corresponding instability in channel flow was found to have no such finite-$Re$ threshold, although for Oldroyd-B fluids, the instability is restricted to ultra dilute solutions with $\beta \gtrsim 0.99$, and very large $Wi=O(10^3)$ \citep{Khalid2021a, Khalid2021b}. Subsequently, these conditions have been relaxed to a more physically-relevant critical $Wi\gtrsim 110$ at $\beta \approx 0.98$, by limiting the maximum extension of the polymers ($L_{max}=70$) in a FENE-P model \citep{buza2021}. 
This suggests that a purely elastic instability can smoothly morph into an elasto-inertial one, where inertia plays a role but the instability is found to only derive its energy through elastic terms. This remains the case even as high as $Re=O(1000)$ \citep{buza2021}. 
While this new instability is active for a wide range of parameter values, it does not appear to overlap with areas where either EIT or ET have been found, consistently appearing at much higher $Wi$ at a given $Re$. Therefore, the question of its relevance to these nonlinear states remains open.

A key issue is whether the branch of travelling wave solutions which emerge from the neutral curve is subcritical and so exist down to some saddle node at Weissenberg number $Wi_{sn}$ below the critical value $Wi_c$, thereby potentially connecting the instability to either EIT and/or ET in parameter space.  
\cite{Page2020} demonstrated the existence of substantial subcriticality albeit at  $Re=60$ (and $\beta=0.9$) where $Wi_{sn}=8.8$ is much lower than $Wi_c=26.7$. Despite EIT not existing at this low $Re$, the upper branch travelling waves found there clearly resembled the `arrowhead' states found in the simulations of \cite{Dubief2020} at $Re=1000$ when EIT was annealed by increasing the elasticity. Weakly nonlinear analysis (in the channel by \cite{buza2021} and pipe flow by \cite{wan2021}) has confirmed the general subcritical nature of the instability but can not give global information about how far $Wi_{sn}(Re,\beta)$ is below $Wi_c(Re,\beta)$.
Our purpose here is to answer this by performing an investigation using branch continuation to track where the travelling waves exist in $(Wi,Re,\beta)$-parameter space. 
This turns out to be feasible, as a branch continuation procedure based on solving an algebraic set of equations derived directly from the governing equations is {\em much} more efficient than branch continuing using a DNS code as done in \cite{Page2020}. There are two reasons for this. Firstly, the travelling wave is highly symmetric: it is 2-dimensional and has a symmetry around the channel's midplane. Secondly, far fewer degrees of freedom are needed to resolve the flow algebraically compared to the number needed to keep a time-stepping code stable. For example, the algebraic formulation needs only $\approx 50$ Chebyshev  modes in the cross-stream direction for convergence at the parameters considered while the DNS code needs $\approx 128$ modes to stay time-stable.

%
%
There have been previous theoretical attempts to generate nonlinear solutions to viscoelastic flow in channels and pipes but without an anchoring bifurcation point. These have centred on constructing a high order expansion assuming the solution is dominantly streamwise and temporally monochromatic and taking the leading state to be one of the least-damped linear modes of the base state \citep{Meulenbroek2003, Morozov2005, Morozov2019}. 

This approach has produced some interesting signs of convergence with an increasing number of terms included in the expansion.
In particular, by taking expansions up to 11th order in the amplitude, \cite{Morozov2005} and \cite{Morozov2019} (plane Couette and channel flow respectively) see apparent convergence  to nontrivial TW solutions in creeping ($Re \ll 1$) flows of upper-convected Maxwell (UCM) fluids ($\beta = 0$) as well as Oldroyd-B fluids at low $\beta$. 
The branch continuation used here is similar in spirit but closer to classical weakly nonlinear theory, and differs in two significant ways: 1) it is firmly rooted in the neutral curve found by \cite{Khalid2021a} -- i.e. the zero amplitude limit smoothly leads to the neutral curve (unknown in \cite{Morozov2019}),
and 2) the order of the expansion is taken as high as necessary (typically 50-80 Fourier modes) to get convergence. 

%
%
The rest of the paper is organised as follows. Section 2 briefly recaps the formulation of viscoelastic channel flow described in our earlier work \citep{buza2021}. Section 3 then outlines the branch continuation approach, with the technical details relegated to a series of Appendices. The results are presented in two sections: section 4 considers finite inertia $Re >0$ and section 5 deals with inertialess flows at $Re=0$. Section 4 exclusively concentrates on $\beta=0.9$ and considers how the subcritical travelling wave branches behave as: 1) $Re$ varies over the range $Re \in [0,3000]$; and 2) as the domain size varies at $Re=30$. 
For the analysis of creeping flow in section 5 at $Re=0$, we explore the existence of the travelling waves over the $(Wi,\beta)$ plane for $\beta \in [0.5,1)$ and $Wi < 50$ and $L_{max} \in\{70,100,500\}$, the maximum polymer extensibility in the FENE-P model. 
\cite{Morozov2022} has concurrently found travelling waves in viscoelastic channel flow at $Re=0.01$ by the complementary approach of  time stepping in the Phan-Thien-Tanner model. 
These waves correspond to the attracting upper branch of the curves shown here. Finally a discussion follows in section 6.

%
%

\section{Formulation}
\label{sect:formulation}

As in \cite{buza2021}, we consider pressure-driven flow of an incompressible viscoelastic fluid in a channel bounded by two parallel, stationary, rigid plates separated by a distance of $2h$.
We model viscoelasticity using the FENE-P model so that the governing equations are 
\begin{subequations} \label{eq:fenep}
\begin{align}
    Re \Big[ \partial_t \bu + \left( \bu \cdot \bnab \right) \bu \Big]+ \bnab p &= \beta \Delta \bu + (1-\beta) \bnab \cdot \T(\C) + \begin{pmatrix}
    F_x \\ 0
    \end{pmatrix}, \label{eq:fenep_u} \\
    \bnab \cdot \bu &=0, \label{eq:fenep_incomp}\\
    \partial_t \C + \left( \bu \cdot \bnab\right) \C + \T(\C) &= \C \cdot \bnab \bu +   \left( \bnab \bu \right)^{T} \cdot \C + \frac{1}{Re Sc} \Delta \otimes \C. \label{eq:fenep_C}
\end{align}
where $(\Delta \otimes \C)_{ij}:= \Delta C_{ij}$.
The constitutive relation for the polymer stress, $\T$, is given by the Peterlin function
\begin{equation}
    \T(\C) := \frac{1}{Wi} \Big( f (\mathrm{tr} \, \C) \C - \mathbf{I} \Big), \quad {\rm where} \quad 
f(s) := \left(1-\frac{s-3}{L^2_{max}}\right)^{-1}
    \label{eq:fenep_constitutive}
\end{equation}
\end{subequations}
with $L_{max}$ denoting the maximum extensibility of polymer chains.
Here $\C \in \mathrm{Pos}(3)$ is the positive definite polymer conformation tensor and
$\beta:= \nu_s/\nu \in [0,1]$ denotes the viscosity ratio where $\nu_s$ and $\nu_p=\nu-\nu_s$ are the solvent and polymer contributions to the total kinematic viscosity $\nu$. The equations are non-dimensionalized by $h$ and the bulk speed  
\begin{equation}
U_b:= \frac{1}{2h} \int^h_{-h} u_x\, dy
\end{equation}
which, through adjusting the imposed pressure gradient $F_x$ appropriately, is kept constant so that the Reynolds and Weissenberg numbers are defined as
\begin{equation}
Re:= \frac{hU_b}{\nu}, \quad Wi:= \frac{\tau U_b}{h}
\end{equation}
where $\tau$ is the polymer relaxation time. 

The Schmidt number $Sc$, appearing solely in the polymer diffusion term and defined as the ratio between the solvent kinematic viscosity and polymer diffusivity \citep{Sid2018}, 
is typically of order $O (10^6)$ in physical applications.
In this work, enhanced diffusion (i.e. lower $Sc$) had to be employed to regularize the hyperbolic equation \eqref{eq:fenep_C}, as is customarily done in other works involving viscoelastic direct numerical simulations \citep{Dubief2020,Sid2018}.
This is also necessitated by the spectral methods embedded in our branch continuation scheme.
 which behave slightly worse than finite difference methods in this respect,
pushing the maximum admissible Schmidt number down to $Sc = 250$ from typically $1000$
\citep[cf.][and Appendix \ref{sect:poldiff}]{Dubief2020}.

Equation \eqref{eq:fenep} is supplemented with non-slip boundary conditions on the velocity field.
For the conformation tensor $\C$, we impose
$$
\partial_t \C + \left( \mathbf u \cdot \bnab\right) \C + \T(\C) = \C \cdot \bnab \bu +   \left( \bnab \bu \right)^{T} \cdot \C + \frac{1}{Re Sc} \partial_{xx} \C
$$
at the wall, i.e.\ we minimize the deviation from the $Sc \to \infty$ limit, where no boundary conditions are necessary.
In the streamwise ($x$) direction, periodic boundary conditions are imposed on both $\bu$ and $\C$. Solutions to \eqref{eq:fenep} of the form
$$
\bvarphi(x,y,t;Wi,Re,\beta) = \bvarphi_b(y;Wi,Re,\beta) + \hat{\bvarphi}(X:=x-ct,y;Wi,Re,\beta),
$$
(where $\bvarphi = (u_X,u_y,p,C_{XX},C_{yy},C_{zz},C_{Xy})$ is the vector composed of all variables) are sought in two consecutive steps. First, the steady, 1-dimensional base state $\bvarphi_b(y;Wi,Re,\beta)$ is solved for numerically at a given $Wi$, $Re$ and $\beta$ (with other model parameters such as $L_{max}$ suppressed for clarity). Then a possibly-large, 2-dimensional perturbation $\hat{\bvarphi}(X,y;Wi,Re,\beta)$ is sought  which is steady in a frame travelling at some {\em a priori} unknown phase speed $c$ in the $\hat{{\bf x}}$ direction.

%
%
\section{Numerical Methods}

For travelling waves (TW), time derivatives can be replaced by $-c\partial_X$ in the governing equations and the problem then becomes elliptic with a `nonlinear' eigenvalue $c$. This approach circumvents the need for time integration but at the price of specialising to steady solutions viewed from some Galilean frame.
Writing the various terms of the governing equations \eqref{eq:fenep} for $\hat{\bvarphi}$ according to their degree of nonlinearity gives
\begin{equation}
\mathcal{L}[\hat{\bvarphi}] + \mathcal{B}[\hat{\bvarphi},\hat{\bvarphi}] + \mathcal{N}[\hat{\bvarphi}] + \F = \mathbf{0},
    \label{eq:operatorform}
\end{equation}
where
$$
\mathcal{L}[\hat{\bvarphi}] := 
\begin{pmatrix}
Re  \Big( -c \partial_X \hat{\bu} + ( \bu_b \cdot \bnab ) \hat{\bu} + ( \hat{\bu} \cdot \bnab) \bu_b \Big) + \bnab \hat{p}  - \beta \Delta \hat{\bu}
\\
\bnab \cdot \hat{\bu}
\\
-c \partial_X \hat{\C} + ( \bu_b \cdot \bnab ) \hat{\C} + ( \hat{\bu} \cdot \bnab) \C_b - 2 \mathrm{sym} \big( \C_b \cdot \bnab \hat{\bu} + \hat{\C} \cdot \bnab \bu_b\big) - \frac{1}{Re Sc} \Delta \otimes \hat{\C}
\\
\end{pmatrix} 
$$
collects the linear contributions,
$$
\mathcal{B}[\hat{\bvarphi}_1,\hat{\bvarphi}_2] := 
\begin{pmatrix}
Re  \, ( \hat{\bu}_1 \cdot \bnab ) \hat{\bu}_2
\\
0
\\
(\hat{\bu}_1 \cdot \bnab) \hat{\C}_2 - 2 \mathrm{sym} \big( \hat{\C}_1 \cdot \bnab \hat{\bu}_2 \big)
\\
\end{pmatrix} 
$$
forms the bilinear part of the nonlinearity, and
$$
\mathcal{N} [ \bvarphi ] := 
\begin{pmatrix}
-(1- \beta) \bnab \cdot  \T (\hat{\C})  
\\
0
\\
\T (\hat{\C}) 
\\
\end{pmatrix},
\qquad \text{and} \qquad
\F := 
\begin{pmatrix}
(1- \beta) \bnab \cdot  \T(\C_b) +\begin{pmatrix}
    F_X \\ 0
    \end{pmatrix}
\\
0
\\
 - \T(\C_b)
\\
\end{pmatrix}
$$
contain the remainder of the terms, with $\mathcal{N}$ representing the general nonlinearity that originates from the constitutive relation 
($\bu_b$ is the base flow and $\C_b$ is the base conformation tensor which, along with the base pressure, make up $\bvarphi_b$). The channel is the 
2-dimensional domain $\Omega = S^1 \times [-1,1]$, with $S^1 = \mathbb{R} / (2 \pi / k)\mathbb{Z}$ denoting a $2 \pi / k$-periodic domain that represents the streamwise ($X$) direction. 

The bifurcating eigenfunction has a symmetry about the midplane -- $(u_X,C_{XX},C_{yy},C_{zz})$ are symmetric in $y$ and $(u_y,C_{Xy})$ are antisymmetric -- which is  preserved at finite amplitude in the subsequent 'arrowhead'-type travelling waves. This is exploited in what follows by only solving the flow in the lower half of the channel $y \in [-1,0]$ and assuming appropriate symmetry conditions at the midplane $y=0$.

\subsection{Branch continuation}

All dependent variables are approximated using a Fourier-Chebyshev basis $\{ \phi_n(X) \psi_m(y) \}_{n,m \in \mathbb{N}}$, where 
\begin{equation}
\phi_n(X):= \sqrt{k/(2 \pi)} \, e^{inkX} \quad {\rm and} \quad
\tpsi_m(y) := \cos [m \cos^{-1}(2y+1)].
\end{equation}
Corresponding to this basis, a TW truncated at order $N_X \times N_y$ may be written as
\begin{equation}
    \hat{\bvarphi}(X,y) =  \sum_{n = - N_X}^{N_X} \sum_{m=0}^{N_y} \ba_{nm}  \phi_n(X) \tpsi_m(y),
    \label{eq:expansion}
\end{equation}
where $\ba_{nm} \in \mathbb{C}^7$ is the vector of coefficients satisfying 
\begin{equation}
    \ba_{(-n)m} = \bar{\ba}_{nm}
    \label{eq:conjugate_coeffs}
\end{equation}
($\bar{\ba}_{nm}$ is the complex conjugate of $\ba_{nm}$). Substituting \eqref{eq:expansion} into \eqref{eq:operatorform}, gives
$$
\sum_{n,m} \mathcal{L}[\ba_{nm} \phi_n \tpsi_m] + 
\sum_{n,m} \sum_{p,q} \mathcal{B} \left[ \ba_{nm} \phi_n \tpsi_m, \ba_{pq} \phi_p \tpsi_q \right] +
\mathcal{N} \left[\sum_{n,m} \ba_{nm} \phi_n \tpsi_m\right] + \F = \mathbf{0}.
$$
A projection onto the $j$-th Fourier mode now yields\footnote{$\mathcal{L}_\ell[\varphi]$ is a slight abuse of notation that stands for $(\mathrm{vec}(\mathcal{L}[\varphi]))_\ell$.}
\begin{multline*}
\sum_{m} \mathcal{L}^j_\ell[\ba_{jm}  \tpsi_m] + 
\sum_{m} \sum_{q} \sum_r \mathcal{B}^{j-r}_\ell \left[ \ba_{rm} \psi_m, \ba_{(j-r)q} \tpsi_q \right] 
+\left\langle \mathcal{N}_\ell \left[\sum_{n,m} \ba_{nm} \phi_n \tpsi_m\right], \phi_j \right\rangle_{L^2(S^1;\mathbb{C})} \\
+ F_\ell \delta_{0j} = 0.
\end{multline*}
where $\mathcal{L}^j$ (and similarly, $\mathcal{B}^j$) is the operator 
$\mathcal{L}$ (and $\mathcal{B}$) modified such that derivatives in the streamwise direction $\partial_X$ are replaced by multiplications with $ikj$.
Thus, the $X:=x-ct$ dependence is now fully eliminated from the equations. 
To treat the $y$ direction, a collocation method is employed over the Gauss-Lobatto points given by
\begin{equation}
    y_s = \frac{1}{2}\left[\cos \left(\frac{s \pi}{N_y} \right)-1\right] \in [-1,0], \qquad s = 0,\ldots,N_y,
    \label{eq:Gauss_L_points}
\end{equation}
Crucially, these are concentrated near {\em both} the channel boundary and the centreline where the resolution is generally most needed. The exception is near the saddle node where the `arrowhead' polymer structure significantly extends into the region between the midplane and boundary of the channel  and is therefore the most challenging to resolve (e.g. see Figure~\ref{fig:Re0main} later).
The resulting system of complex algebraic equations are 
\begin{align}
\sum_{m} \mathcal{L}^j_\ell[\ba_{jm}  \psi_m](y_s) &+ 
\sum_{m} \sum_{q} \sum_r \mathcal{B}^{j-r}_\ell \left[ \ba_{rm} \tpsi_m, \ba_{(j-r)q} \tpsi_q \right](y_s) \nonumber \\
&+ \left\langle \mathcal{N}_\ell \left[\sum_{n,m} \ba_{nm} \phi_n \tpsi_m \right](y_s), \phi_j \right\rangle_{L^2(S^1;\mathbb{C})} 
+ F_\ell (y_s) \delta_{0j}= 0, \nonumber \\
&\text{for} \quad j = 0,\ldots,N_X, \quad s = 0,\ldots,N_y, \quad \ell = 1,\ldots,7,
    \label{eq:main_anm}
\end{align}
for the coefficients $\ba_{nm} \in \mathbb{C}^7$ with $n,m \geq 0$.
The remainder of the coefficients in \eqref{eq:expansion} are computed via \eqref{eq:conjugate_coeffs}.

Two further equations are needed to determine the wave speed $c$ and the applied pressure gradient  $F_X$. As indicated above, $F_X$ is determined by ensuring the perturbation volume flux vanishes,
\begin{equation}
    \int^1_{-1}\hat{u}_X dy=0,
    \label{eq:bonuseq1}
\end{equation}
and 
\begin{equation}
    \mathrm{Im} \int^{2\pi/k}_0 e^{-ikX} \hat{u}_X(X,y_{15}) \, dx =0
    \label{eq:bonuseq2}
\end{equation}
is imposed to eliminate the phase degeneracy of the travelling wave and thereby determine the wave speed (the exact collocation point $y_{15}$ is chosen arbitrarily, e.g. see \cite{wedin2004}).
The resulting nonlinear, complex, algebraic system of equations comprising \eqref{eq:main_anm}, \eqref{eq:bonuseq1} and \eqref{eq:bonuseq2} reads
\begin{equation}
    \mathcal{F} (\ba,c,F_X;k,Wi,Re,\beta,Sc) = \mathbf{0}, 
    \label{eq:Feq}
\end{equation}
where $\ba = \mathrm{vec}((a_{nm})_\ell)$.
System \eqref{eq:Feq} gives rise to $Q:=2+2 \times 7 \times N_X \times (N_y+1) + 7 \times (N_y+1) \sim 14N_X N_y$ real nonlinear equations to be solved simultaneously (by slight abuse of notation we shall denote the real parts of $\mathcal{F}$ and $\ba$ in \eqref{eq:Feq} by the same letters in what follows).
Steady states of interest may now be extracted from \eqref{eq:Feq} using a Newton-Raphson root finding scheme given a good enough initial guess. The neutral curve found by \cite{Khalid2021a} and the the weakly nonlinear analysis in \cite{buza2021} are used to generate this initially. Then pseudo-arclength continuation -- see Appendix \ref{A} -- is used to proceed along the solution branch to higher amplitudes away from the neutral curve.

Simulations were typically run at $(N_X,N_y) = (50,60)$ where $Q \approx 43,000$ real degrees of freedom or $(N_X,N_y) = (40,50)$ ($Q \approx 29,000$), depending on the complexity of the tracked states, with occasional grid-convergence checks at much higher resolutions up to $(N_X,N_y) = (80,80)$ ($Q \approx 91,000$); see Appendix B.
Generally, lower branch solutions were less resolution-dependent, and required about half the Fourier modes of their upper branch counterparts. The minimum requirement for the number of Chebyshev modes, $N_y$, was around $40$ across all parameter regimes, with a slight increase to $50$ around saddle-node points due to the suboptimal placement of collocation points for this region. Reducing the polymer diffusion increases the requirements both in $N_X$ and $N_y$, and adjustments in $k$, the domain size, necessitate equivalent adjustments in $N_X$.

\subsection{Direct numerical simulations} \label{sec:dns_appendix}

The Dedalus codebase \citep{burns2020dedalus} was used to time step eqs.~\eqref{eq:fenep} in order to examine the stability of the TWs found. To allow this DNS code to interface seamlessly with the branch continuation code, the simulations were also performed on the half-channel using exactly the same symmetry boundary conditions described above and using the same spectral expansions. This allowed an unstable lower branch solution of the branch continuation procedure to be used directly as an initial condition for the DNS and the fact that this remained steady under time-stepping provided a valuable cross-check of the two approaches. 

In the DNS, the full state $\bvarphi$ was minimally expanded into $N_x=128$ Fourier modes in the periodic $x$-direction and into $N_y=128$ Chebyshev modes in the wall-normal direction with higher resolutions of 256 and 512 available in either or both dimensions to check truncation robustness. The equations were advanced in time using a 3rd-order semi-implicit BDF scheme \citep{wang2008variable} and a constant timestep $\Delta t=5\times 10^{-3}$. 

%
%
%
\section{Results: Travelling waves at finite $Re$ for $(\beta, L_{max},Sc) = (0.9,500,250)$}

As in \citet{Page2020} and \citet{buza2021}, we fix $\beta = 0.9$ and $L_{max} = 500$ for the initial set of computations.
The Schmidt number had to be chosen slightly smaller than that of \citet{Page2020} and \citet{Dubief2020} at $Sc = 250$ due to the considerations given in Section \ref{sect:formulation} and Appendix \ref{sect:poldiff}.

%
%
\begin{figure}
\includegraphics[width=1\textwidth]{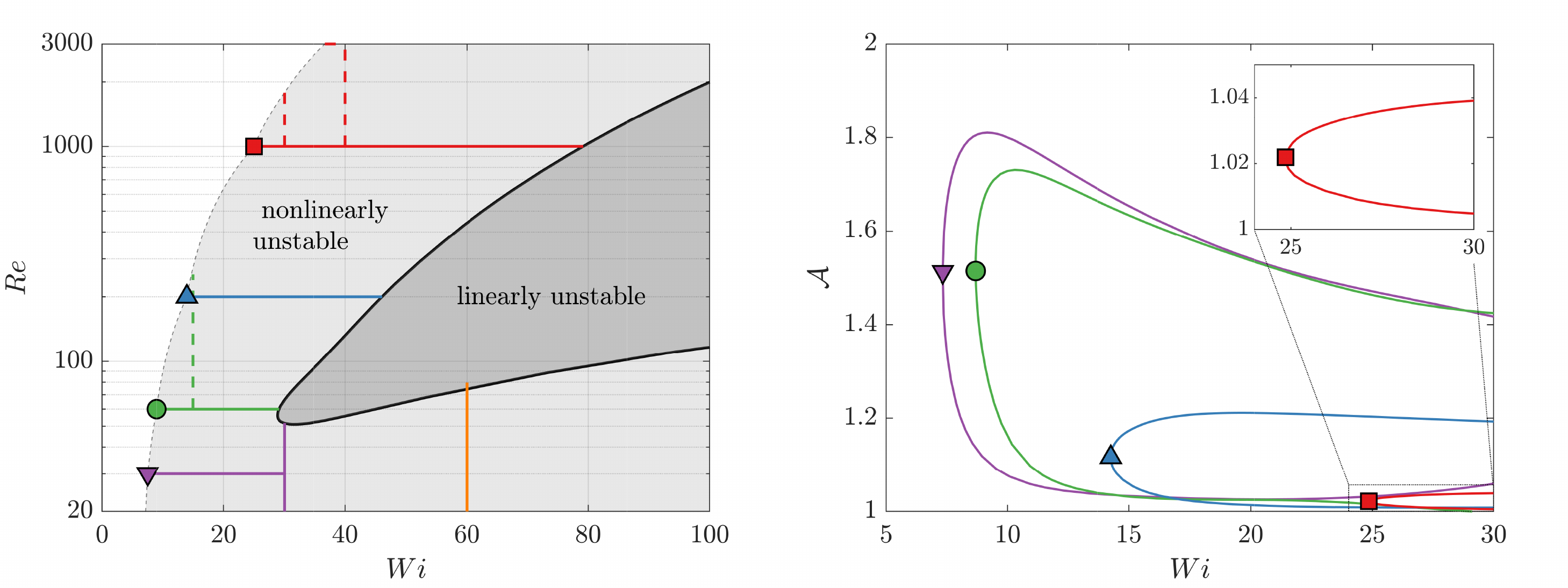}
\centering
\caption{(Left) Linearly and nonlinearly unstable regions in the $Wi-Re$ plane for $\beta=0.9$, $L_{max}=500$ and $Sc=250$. The saddle node Weissenberg numbers $Wi_{sn}(Re,k)$ shown are: $Wi_{sn}(30,1.6)=7.6$, $Wi_{sn}(60,1.8)=8.7$, $Wi_{sn}(200,2.7)=14.3$, $Wi_{sn}(1000,4.7)=24.9$ and $Wi_{sn}(3000,4.7)=36.7$. Coloured horizontal lines correspond to branches on the right panel and symbols indicate the saddle-node points. (Right) Solution branches tracking travelling waves as $Wi$ varies at constant $Re \in \{30,60,200,1000\}$ (note horizontal axis is $Re=20$). 
}
\label{fig:uno}
\end{figure}

Upon supplying the weakly nonlinear predictions as initial conditions to the continuation routine, any branch of solutions emanating from the neutral curve can be tracked  starting directly from its bifurcation point.
Three branches were launched downwards in $Wi$ at fixed $Re = 1000,200,60$, starting from their respective bifurcation points at $k_{opt} = 4.7, 2.7, 1.8$.
These wave numbers are optimal in the sense of marginal stability and so do not necessarily minimise $Wi_{sn}(Re,\beta)$, but do provide a good upper estimate of it.
An additional, fourth branch was initialized from the lowest point on the neutral curve at $Wi = 30$ and $k_{opt} = 1.6$, continued down to $Re = 30$ at fixed $Wi$, then -- after a switch in direction -- towards decreasing $Wi$ at fixed $Re$.
A schematic depiction of these branches is given in the left panel of Figure~\ref{fig:uno}.\\

%
%
\begin{figure}
\includegraphics[width=1\textwidth]{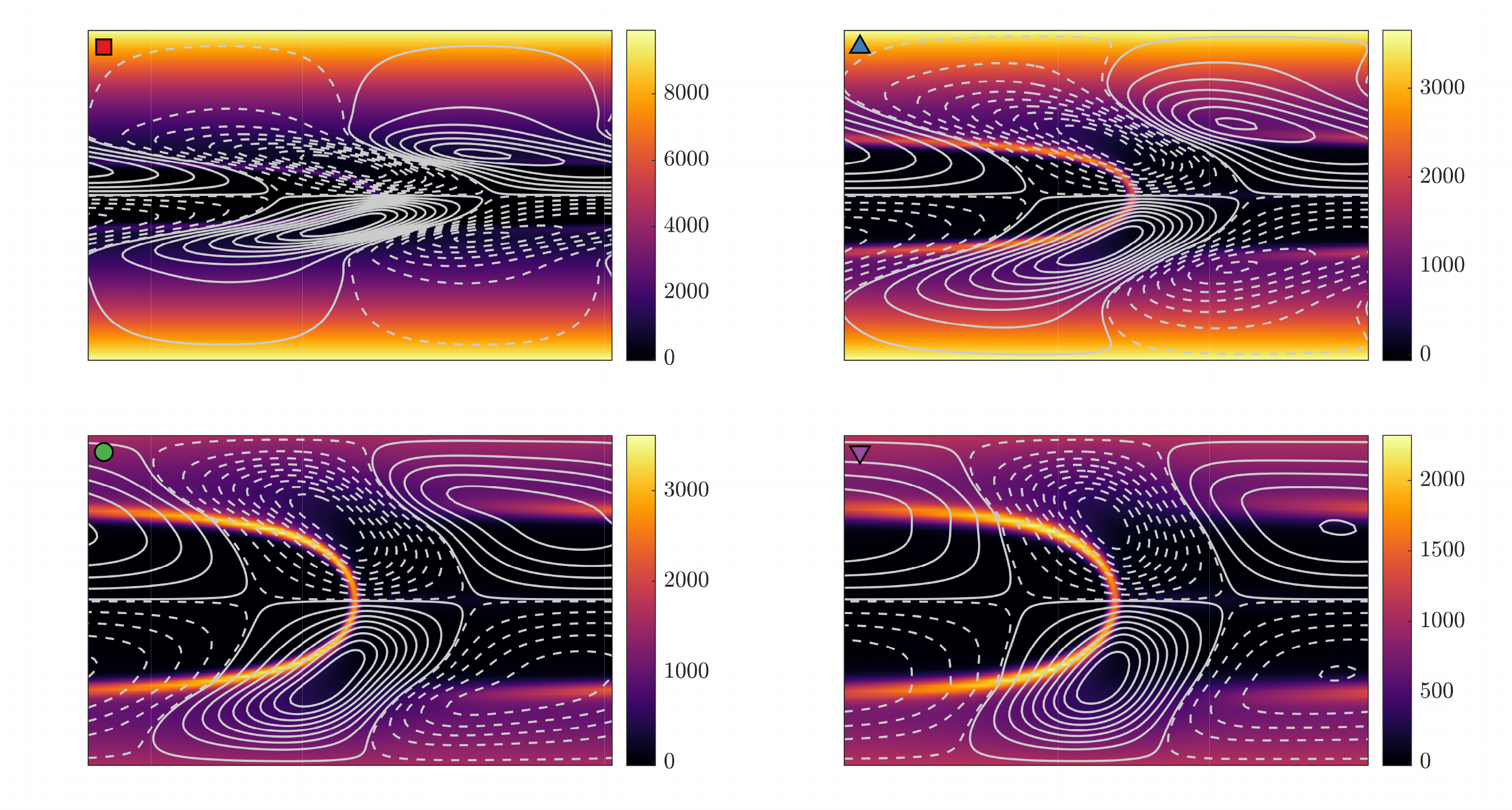}
\centering
\caption{Snapshots of full states, in terms of $\mathrm{tr} \, \C$ (contours), at the saddle-node bifurcation points from Figure~\ref{fig:uno}. Lines correspond to level sets of the perturbation stream function. 
The noticeable thinning of the arrowhead structure with increasing $Re$ (from bottom right to top left) is partially due to the corresponding increase in $k_{opt}$ (and decrease in domain length; see Section~\ref{sect:domainlength} and compare with Figure~\ref{fig:k}). The domain size is $2 \pi / k_{opt}$ in the $X$ direction, with $k_{opt} \in \{4.7,2.7,1.8,1.6\}$ in panel order.
}
\label{fig:saddles}
\end{figure}

The right panel of Figure~\ref{fig:uno} shows the amplitude evolution of these four branches.
As a measure of amplitude, we chose the volume-averaged trace of the polymer conformation relative to the laminar value, i.e.,
\begin{equation}
    \mathcal{A}: = \frac{\langle \mathrm{tr} \, \C \rangle_\Omega}{\langle \mathrm{tr} \, \C_b \rangle_\Omega},
    \label{eq:ampdef}
\end{equation} 
again, to remain consistent with \citet{Page2020}. The lower $Re$ branches of Figure~\ref{fig:uno} (right) are reminiscent of the branch shown in \citet{Page2020}\footnote{In fact, the green branch is at the same Reynolds number ($Re = 60$), albeit with different $k$ and $Sc$.}, and the higher $Re$ ones are lower amplitude  variants of these.
This shrink in relative amplitude can be attributed to the increase in both $Re$ and $k$, with the latter playing a non-negligible role through the accompanying change in domain size (see Section~\ref{sect:domainlength}).\\

\begin{figure}
\includegraphics[width=1.05\textwidth]{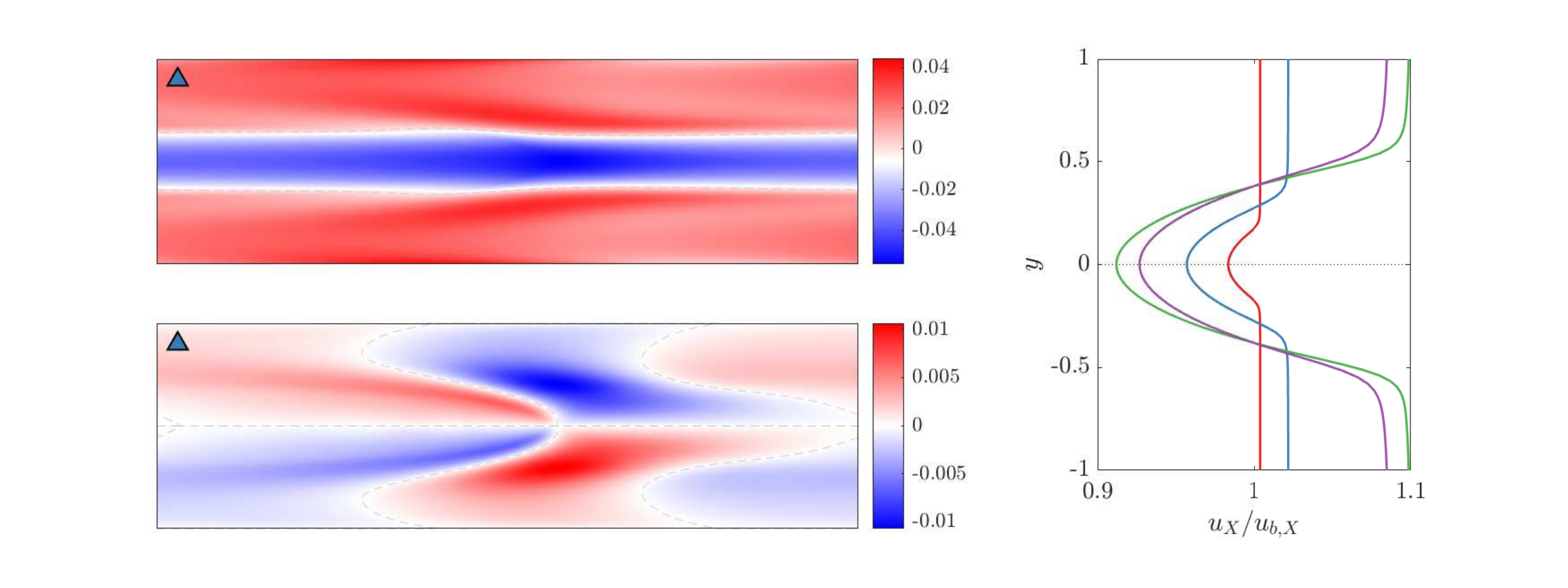}
\centering
\caption{(Top left) Contours of $\hat{u}_X/u_{b,X}$ at the saddle-node point of the $Re = 200$ branch (blue triangle in Figures~\ref{fig:uno}~and~\ref{fig:saddles}). The domain length is $2 \pi / 2.7$ as in Figure~\ref{fig:saddles}.
(Bottom left) Same with $\hat{u}_y/u_{b,X}$ displayed.
(Right) Mean velocity profiles $u_X/u_{b,X}$ at the saddle-node points marked on Figures~\ref{fig:uno}~and~\ref{fig:saddles}, with color codes matching that of Figure~\ref{fig:uno}.
}
\label{fig:questionmark}
\end{figure}

We explore one of these states (the saddle node from the $Re=200$ branch) further in figure \ref{fig:questionmark}, where we report the perturbation velocities as a fraction of the local base streamwise velocity, $u_{b,X}$.
The arrowhead of polymer stretch close to the centreline is associated with a backwards `jet' in the perturbation streamwise velocity. 
The contours of vertical velocity indicate a change in sign of $\partial_y \hat{u}_y$ across a stagnation point, which is consistent with the physical interpretation of the self-sustaining mechanism proposed by \citet{Morozov2022}.

%
%

\subsection{General interpretation of solution branches} 
\label{sect:branches}

%
%
\begin{figure}
\includegraphics[width=1.05\textwidth]{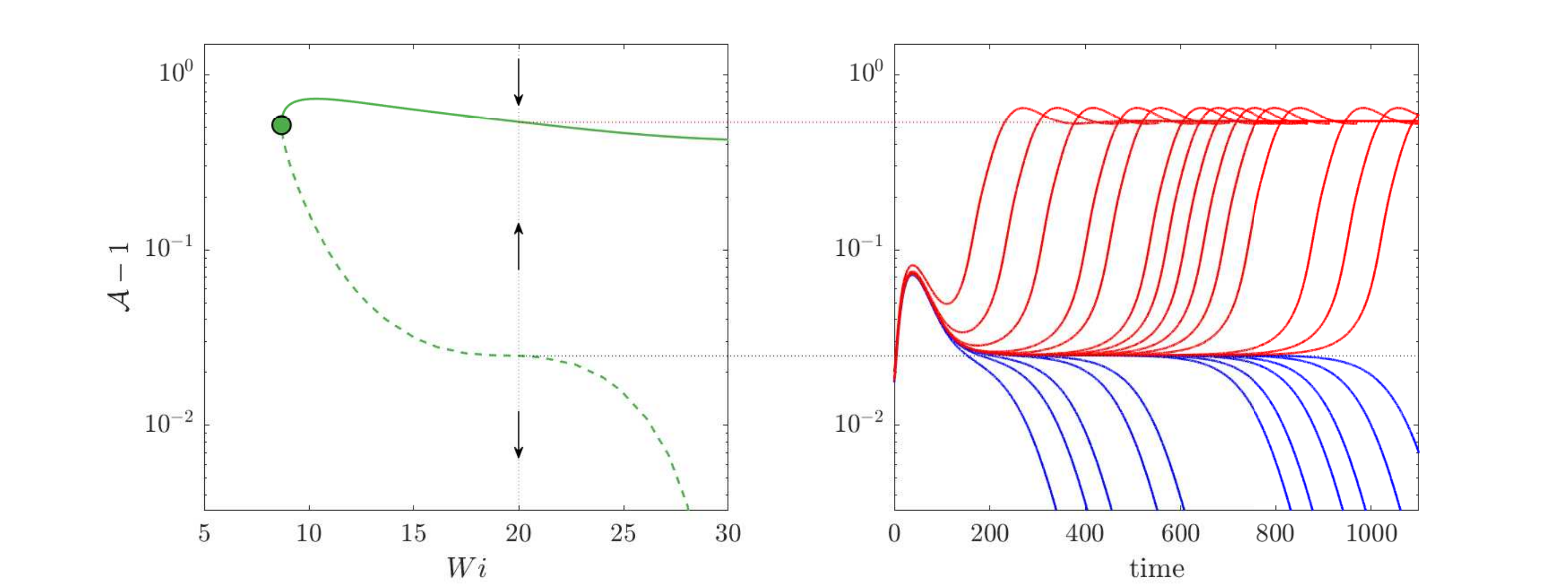}
\centering
\caption{(Left) The $Re = 60$ branch from Figure~\ref{fig:uno}, with edge states indicated by the dashed line. (Right) Results of the independent edge-tracking algorithm at Wi=20. The blue/red lines indicate evolutions which start close to the lower branch state and converge to the laminar/upper branch state respectively.}
\label{fig:edge}
\end{figure}

Qualitatively, all solution branches behave in a similar way.
A sample case is depicted in Figure~\ref{fig:edge}, showcasing the main features.
The lower branches emanating from the neutral curve are all unstable until reaching their respective saddle-node  points, labeled by a variety of symbols in Figure~\ref{fig:uno} (circle for our sample branch in Figure~\ref{fig:edge}), with the corresponding states shown in Figure~\ref{fig:saddles}. Points on the (unstable) lower branches are found to be edge states which are attracting states on a codimension-1 manifold separating two different basins of attraction \citep{skufca2006edge,schneider2008laminar,duguet2008transition}. This is illustrated by Figure~\ref{fig:edge} (right) at $Wi=20$ and $Re=60$ where an edge-tracking procedure, applied between the upper branch and laminar states, converges on the lower branch state. The lower branch state is a saddle  but with only one unstable direction either pointing to the laminar or upper branch state. Upper branch states, at least $Re = 60$ \citep{Page2020}, start as stable nodes as $Wi$ increases away from $Wi_{sn}$ but quickly experience Hopf bifurcations to tertiary states (if the base state is the `primary'). These bifurcations and where these tertiary states lead are interesting questions beyond the scope of this manuscript.

Based on the above observations, we have the following picture: if the laminar state is disturbed with a perturbation large enough to reach a certain threshold, determined by the minimal amplitude of approach of the stable manifold of a lower branch state, the flow will evolve towards the upper branch, forming a stable travelling wave. The threshold amplitude to trigger growth is bounded above by the amplitude of the lower branch state itself, which remains $\mathcal{A} < 1.05$ across the domain of existence of travelling waves. In other words, this domain (shaded bright grey on Figure~\ref{fig:uno}) is nonlinearly unstable when subjected to finite but small amplitude disturbances.
\subsection{Influence of domain length}
\label{sect:domainlength}

%
%
\begin{figure}
\includegraphics[width=1\textwidth]{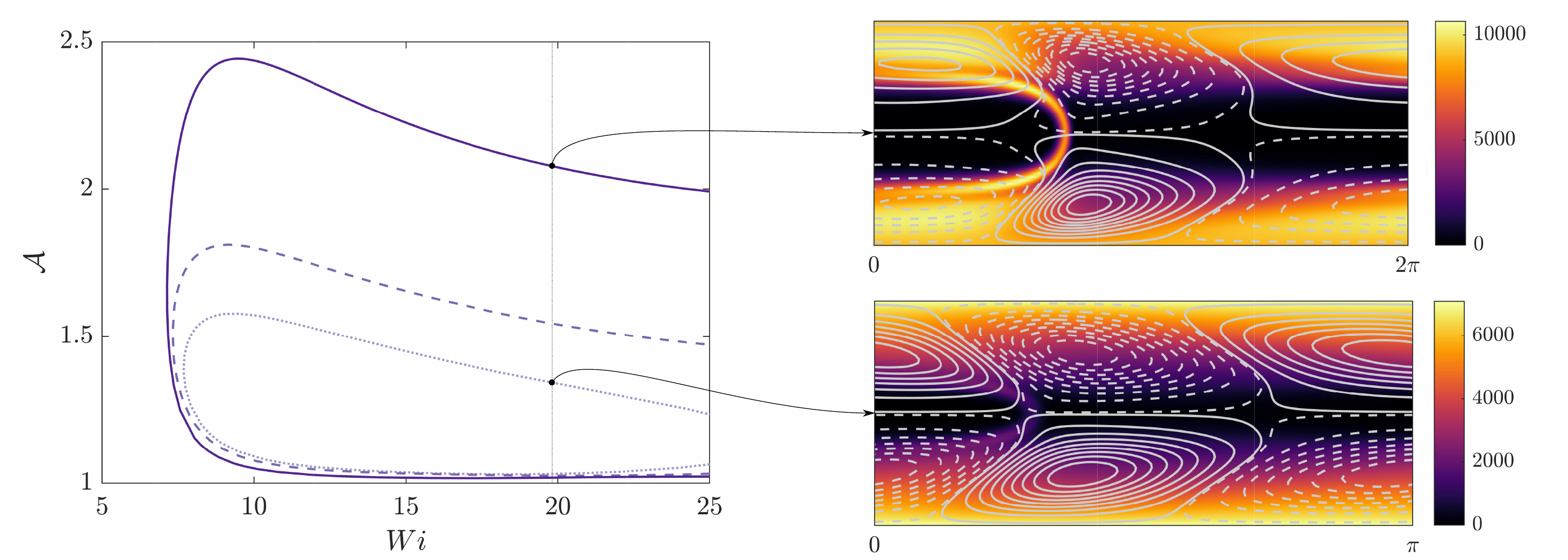}
\centering
\caption{(Left) Branches of travelling waves at different wave numbers - and thus domain sizes - at $k = 1$ (\sampleline{}), $k$ $(= k_{opt})= 1.6$ (\sampleline{dashed}) and $k = 2$ (\sampleline{dotted}) with fixed $Re=30$ ($k=1.6$ is also shown colored purple in Figure~\ref{fig:uno}).
(Right) Snapshots of $k = 1$ and $k = 2$ upper branch travelling waves at fixed $Wi= 19.814$ (note the difference in domain size).
Contours correspond to $\mathrm{tr} \, \C$ and lines correspond to level sets of the perturbation stream function. All visible differences are contained in the superimposed wave solutions - the laminar state does not depend on $k$.}
\label{fig:k}
\end{figure}

This section is dedicated to studying the effect of $k$, and thus the influence of domain size $[0,2 \pi/k]$ on the TWs.
Figure~\ref{fig:k} shows how a single branch of travelling waves at $Re = 30$ (purple in Figure~\ref{fig:uno})  changes with $k$.
It has already been established that the steady arrowhead structure is highly sensitive to domain length in the EIT regime \citep{Dubief2020}. There, through capturing larger scale motions, an increase in domain length was found to unveil structures of increasing complexity, with the possibility of inducing chaotic dynamics at certain parameter combinations.
Similar tendencies can be observed in our case (cf.\ Figure~\ref{fig:k}): An increase in $k$ (and thus decrease in domain length) 
has a considerable weakening effect on the arrowhead structure,
eventually resulting in a complete eradication of TWs
and a subsequent relaminarization. Despite this observation, the location of the saddle-node points seems largely unaffected by  $k$  (cf.\ Figure~\ref{fig:k}), making the marked 'nonlinearly unstable' region on Figure~\ref{fig:uno} robust to changes in the assumed periodicity and domain size. 

\subsection{High elasticity regime: $Re \rightarrow 0$}
\label{sect:highelast}

The high elasticity regime is difficult to access using time-stepping as it becomes increasing stiff as $Re \rightarrow 0$. The algebraic approach taken here suffers no such problems and we can approach and even consider $Re=0$ (see the next section) without difficulty. 

The existence of the centre-mode linear instability at $Re = 0$ is already known in the limit of very dilute polymer solutions ($\beta \to 1$) for $Wi = O (10^3)$ in Oldroyd-B fluids \citep{Khalid2021b} and for $Wi = O (10^2)$ in FENE-P fluids at finite extensibility ($L_{max}$) \citep{buza2021}. To substantiate its connection to ET, the time evolution of these growing modes has to be tracked to see whether they are able to produce turbulent behaviour, presumably after
transitioning through a cascade of intermediate states. Our goal here is to see where the first level of intermediate state - the TWs - exist at low and vanishing $Re$. 

\begin{figure}
\includegraphics[width=0.75\textwidth]{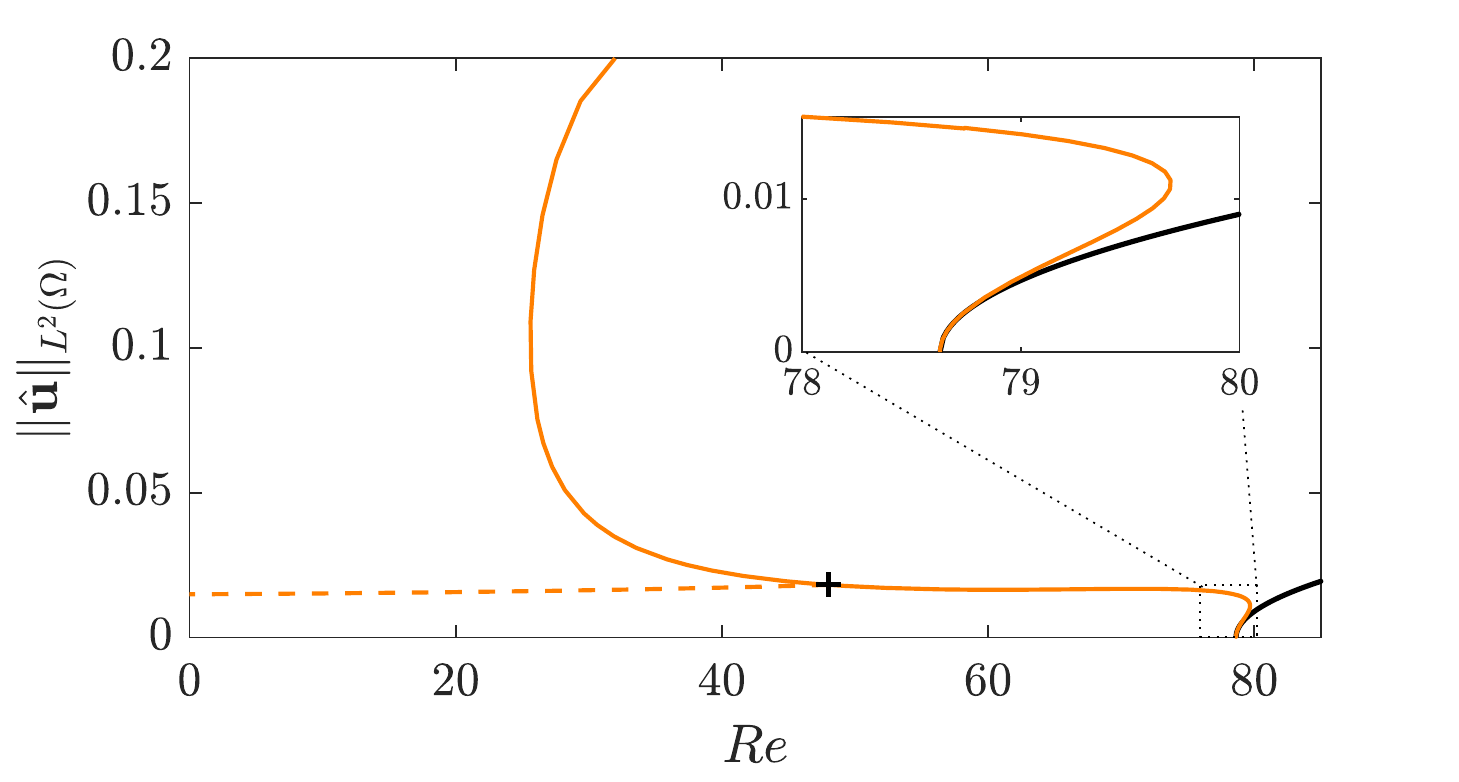}
\centering
\caption{Solution branch at fixed $Wi = 60$, indicated by orange on Figure~\ref{fig:uno} ($k = 1$). The solid black line shows the weakly nonlinear prediction of supercriticality. In orange are the results from branch continuation with the $1/(ReSc)$ formulation (\sampleline{}) and with the $\lambda$ formulation (\sampleline{dashed}) ($\lambda = 0.005$ and $Sc = 250$) which shows that the branch of TWs quickly turns around and heads to lower $Re$ i.e. the TWs are substantially subcritical.}
\label{fig:Wi60branch}
\end{figure}

Weakly nonlinear theory predicts supercritical behaviour in the high elasticity ($Wi/Re$) regime, i.e., along the lower boundary of the linearly unstable region. 
To probe this,  a fifth branch was initiated at fixed $Wi = 60$, starting upwards in $Re$ as indicated by the weakly nonlinear analysis \citep{buza2021} from a marginally stable point in this region (indicated by orange on Figure~\ref{fig:uno}). The resulting branch of solutions is shown in Figure~\ref{fig:Wi60branch}. Given the supercriticality, this branch starts off as a stable node, moving up in $Re$. Almost immediately after leaving the initial bifurcation point (of linear stability), it reaches a saddle-node bifurcation point, turns around and proceeds to advance towards decreasing $Re$, maintaining a relatively low amplitude until reaching a second saddle-node and transitioning to the upper branch. This is an example of how local information provided by weakly nonlinear analysis can be misleading. In fact, the neutral curve gives rise to TWs which reach to lower $Wi$ at fixed $Re$ and lower $Re$ at fixed $Wi$ as shown by Figure \ref{fig:uno}.

%
%

%
%
Upon further inspection, it turns out that the lower (secondary) fold shown in Figure \ref{fig:Wi60branch} at $Wi \approx 30$ is purely a feature of the polymer diffusion term $1/(ReSc) \Delta \otimes C$ growing artificially large (as $Re$ is decreased), the effect of which is already known to destroy small scale dynamics \citep{Dubief2020}. It turns out that the point at which the saddle-node bifurcation occurs can be delayed arbitrarily by adjusting $Sc$ in accordance with the variations in $Re$ to keep the polymer diffusion finite. Numerical experimentation suggested a revised polymer diffusion term of the form
$$
\frac{\lambda}{Wi} \Delta \otimes \C,
$$
for some fixed number $\lambda$. The choice of an inverse scaling with $Wi$ is motivated by observations at $Re = 0$ shown in Appendix~\ref{sect:poldiff}.
If $\lambda = 0.005$ is enforced for the branch in question, which amounts to fixing the coefficient $1/(ReSc)$ at the point marked by '$+$' in Figure~\ref{fig:Wi60branch}, the branch of solutions can be followed down to $Re = 0$ along the lower branch (cf.\ the dashed line in Figure~\ref{fig:Wi60branch}).

%
%
\section{Results: Travelling waves in the creeping flow limit $Re = 0$}
\label{sect:Re0}

%
%
\begin{figure}
\includegraphics[width=1\textwidth]{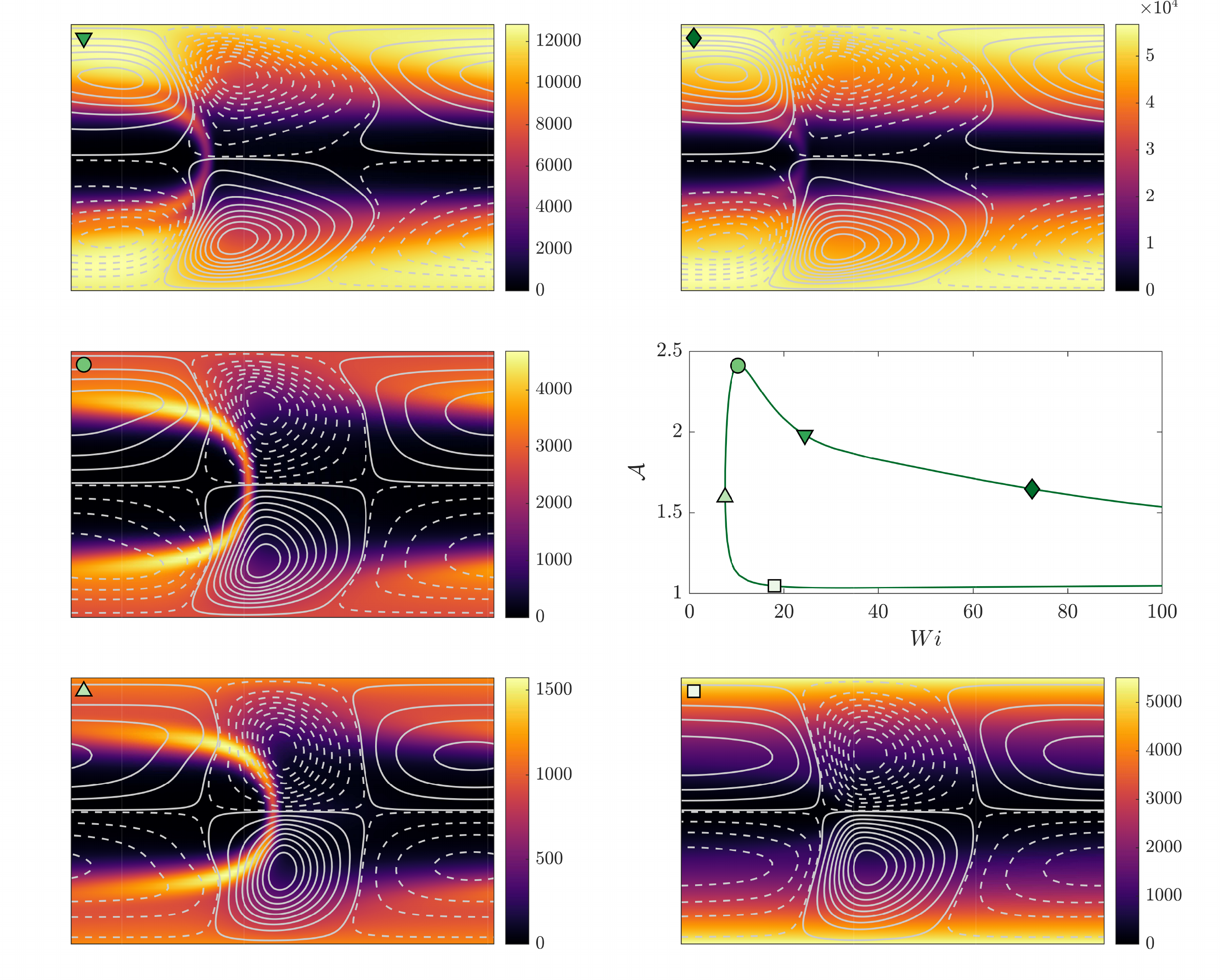}
\centering
\caption{(Middle right) Branch of travelling waves at $Re = 0$, $\beta = 0.9$, $L_{max} = 500$, $\lambda = 0.005$, $k=1$. All other panels correspond to states at the locations marked via different symbols. In these plots, contours correspond to $\mathrm{tr} \, \C$ and lines correspond to level sets of the perturbation stream function. Domain length is $2 \pi$ in all state plots.}
\label{fig:Re0main}
\end{figure}

Once $Re = 0$ is reached, we redirect the continuation tool towards decreasing $Wi$.
The resulting branch takes the familiar shape (from the $Re>0$ cases) shown in Figure~\ref{fig:Re0main}, attaining its saddle-node bifurcation point at $Wi \approx 7.5$, which serves as a lower bound for the region where travelling waves exist \citep[note the waves found by][at $Re=0$ are all above $Wi=20$, albeit with a different model)]{Morozov2022}.
Figure~\ref{fig:Re0main} gives a detailed description of this branch, containing snapshots of full states that illustrate how these waves evolve as $Wi$ is varied. Arrowhead-shaped structures are still clearly visible at low $Wi$ (cf.\ panels on the left side of Figure~\ref{fig:Re0main}), establishing their prevalence even in the high elasticity regime.

For this particular case, specify $k=1$, $L_{max}=500$, $\lambda=0.005$), the stability of steady states was examined along the upper branch using DNS.
At four points, $Wi = 10,20,30,50$, solutions of the branch continuation tool were transferred into the Dedalus based DNS code, and were subsequently subjected to disturbances of finite amplitude. The perturbations were constructed from snapshots extracted from separate simulations of EIT at high-$Re$, which were pre-multiplied by $10^{-6}$ and added to the travelling waves.
All perturbed states returned to their respective stable upper-branch solutions after a period of transient growth, suggesting that two-dimensional ET cannot be initiated from these travelling waves in a direct manner.

\subsection{$\beta \in [0.5,1)$: Relation to recent experiments}

The first experiments claiming to see nonlinear instability in viscoelastic channel flow were performed by Arratia and colleagues \citep{Pan2013,qin2017,Qin2019}. Finite amplitude perturbations were induced by an array of obstacles placed upstream, with the number of obstacles serving as a measure of amplitude. Based on measurements taken further downstream, far away from the initial disturbances, they conclude the existence of a subcritical nonlinear instability that persists down to $Wi \approx 5.4$. With the caveats that their channel had  a square cross-section and FENE-P is an approximation, their results are encouragingly comparable to the 2D channel prediction made here of $Wi_{sn}=7.5$. In Figure S1 of the supplementary material to \citet{Pan2013}, the authors indicate the boundary to the observed instability in a $Wi$ vs.\ perturbation amplitude plane, essentially matching our predictions for the threshold of nonlinear instability given by the $Re = 0$ lower branch (shown in the middle-right panel of Figure~\ref{fig:Re0main})\footnote{
That the unstable region in our case is bounded by the branch is an immediate consequence of the discussion in Section~\ref{sect:branches}: Once the threshold amplitude of a lower branch edge state is reached, solutions continue to grow.
}.
However, in later proceedings, the authors claim that the unstable flow remains time dependent with features reminiscent of ET \citep{qin2017,Qin2019}, as opposed to the upper branch travelling wave scenario described here.

%
%
\begin{figure}
\includegraphics[width=1\textwidth]{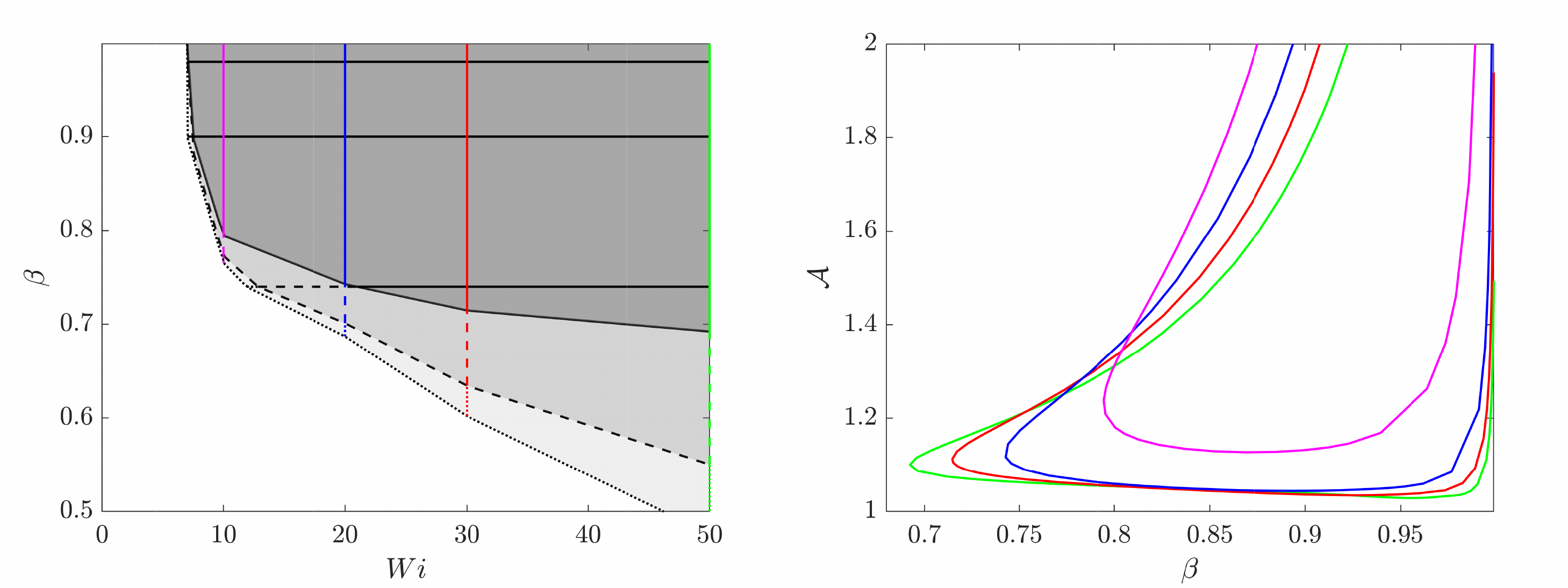} 
\centering
\caption{(Left) Nonlinearly unstable regions in the $Wi-\beta$ plane at $Re = 0$ for $\lambda = 0.005$ (\sampleline{}), $\lambda = 0.003$ (\sampleline{dashed}), $\lambda = 0.002$ (\sampleline{dotted}). Horizontal and vertical lines correspond to solution branches computed via the continuation routine. (Right) Branches with respect to $\beta$, obeying the same color code as on the left panel (at $\lambda = 0.005$).}
\label{fig:Wibeta}
\end{figure}

More recently, Steinberg and coworkers \citep{jha2020,Shnapp2021} have
obtained results in an experimental setup using a channel with a width/height ratio of 7 and so more approximately 2D. However, the viscosity ratio was  $\beta = 0.74$, significantly smaller than the above presented $\beta = 0.9$.
Other recent experiments in a pipe have also considered smaller $\beta$ \cite[e.g.][at $\beta = 0.56$]{Choueiri2021}. Motivated by this, we also performed a few TW branch continuations (at $Wi = 10, 20, 30$ and $50$) with decreasing $\beta$ in an attempt to map out the nonlinearly-unstable domain in the $Wi-\beta$ plane (all with zero inertia).
Results from these computations are shown in Figure~\ref{fig:Wibeta}, with the left panel indicating the unstable region and the right panel containing the solution branches found with varying $\beta$. It transpires that at lower $\beta$, the solutions are a little more sensitive to the artificial diffusion (see Appendix~\ref{sect:poldiff} for further details), necessitating multiple simulations at different $\lambda$ values, all of which are also shown in Figure~\ref{fig:Wibeta}.

The results of \citet{jha2020} and \citet{Shnapp2021} indicate the existence of `elastic waves' over a large range of Weissenberg numbers at $Re \approx 0$ and $\beta = 0.74$. Figure~\ref{fig:Wibeta} is consistent with these  observations for $Wi \geq 11$ at the very least $\beta =0.74$ branch.
Moreover, the latter work \citep{Shnapp2021} implies that for $Wi$ large enough (above $Wi_c = 125 \pm 25$ according to \citet{Shnapp2021}), arbitrarily small perturbations are sufficient to trigger these growing, elastic waves.

%
%
\begin{figure}
\includegraphics[width=0.75\textwidth]{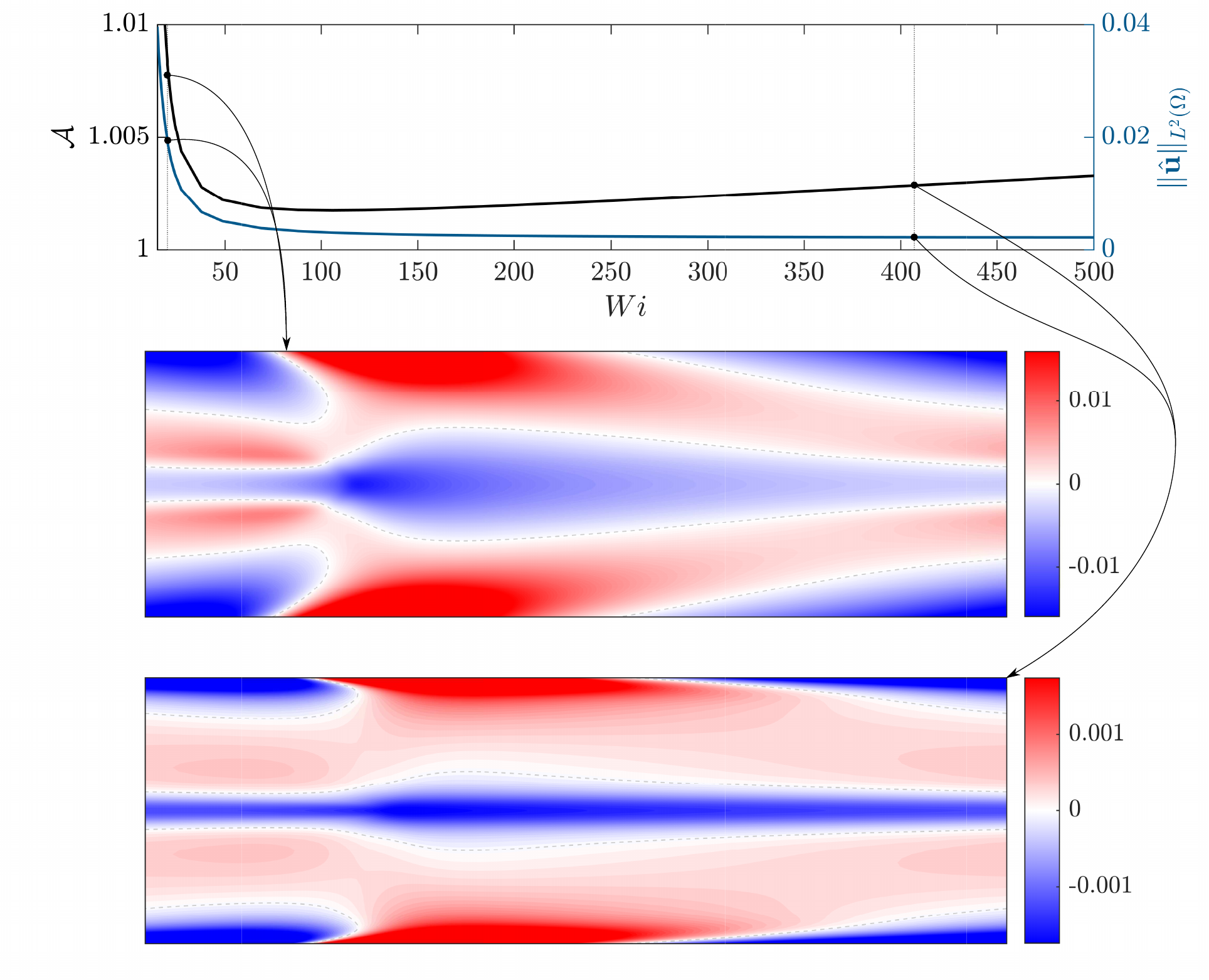}
\centering
\caption{(Top) The $\beta = 0.74$ lower branch from Figure~\ref{fig:Wibeta}, at $\lambda = 0.001$, with respect to the standard amplitude measure $\mathcal{A}$ (in black) and the perturbation velocity magnitude $\Vert \hat{\bu} \Vert_{L^2 (\Omega)}$ (in blue). 
(Middle) State snapshot at $Wi = 20$ from the branch above. Contours show $\hat{u}_X/u_{b,X}$.
(Bottom) State snapshot at $Wi = 407$ from the branch above. Contours show $\hat{u}_X/u_{b,X}$.}
\label{fig:Au}
\end{figure}

In an attempt to recover these observations in the present setting, Figure~\ref{fig:Au} tracks changes in the lower branch TW amplitude, which is an upper bound on the threshold for growth, as $Wi$ is increased.
For the parameter combination in question ($\beta = 0.74$, $Re = 0$), both our standard amplitude measure $\mathcal{A}$ and a separate measure for the magnitude of velocity perturbations, $\Vert \hat{\bu} \Vert_{L^2 (\Omega)}$, are shown.
For $Wi > 30$, $\mathcal{A}$ remains below $1.005$, implying that perturbations amounting to $0.5\%$ of the laminar conformation field $\mathrm{tr} \, \C_b$ are sufficient to trigger growth, making this scenario
practically indistinguishable from a linearly unstable one.
$\mathcal{A}$ reaches its minimum roughly around $Wi = 100$, then -- despite setting off in an increasing trend -- remains negligibly small for the remainder. Perhaps more in line with experimental results, the minimal disturbance momentum $\hat{\bu}$ decreases steadily with respect to $Wi$.

%
%
\begin{figure}
\includegraphics[width=0.6\textwidth]{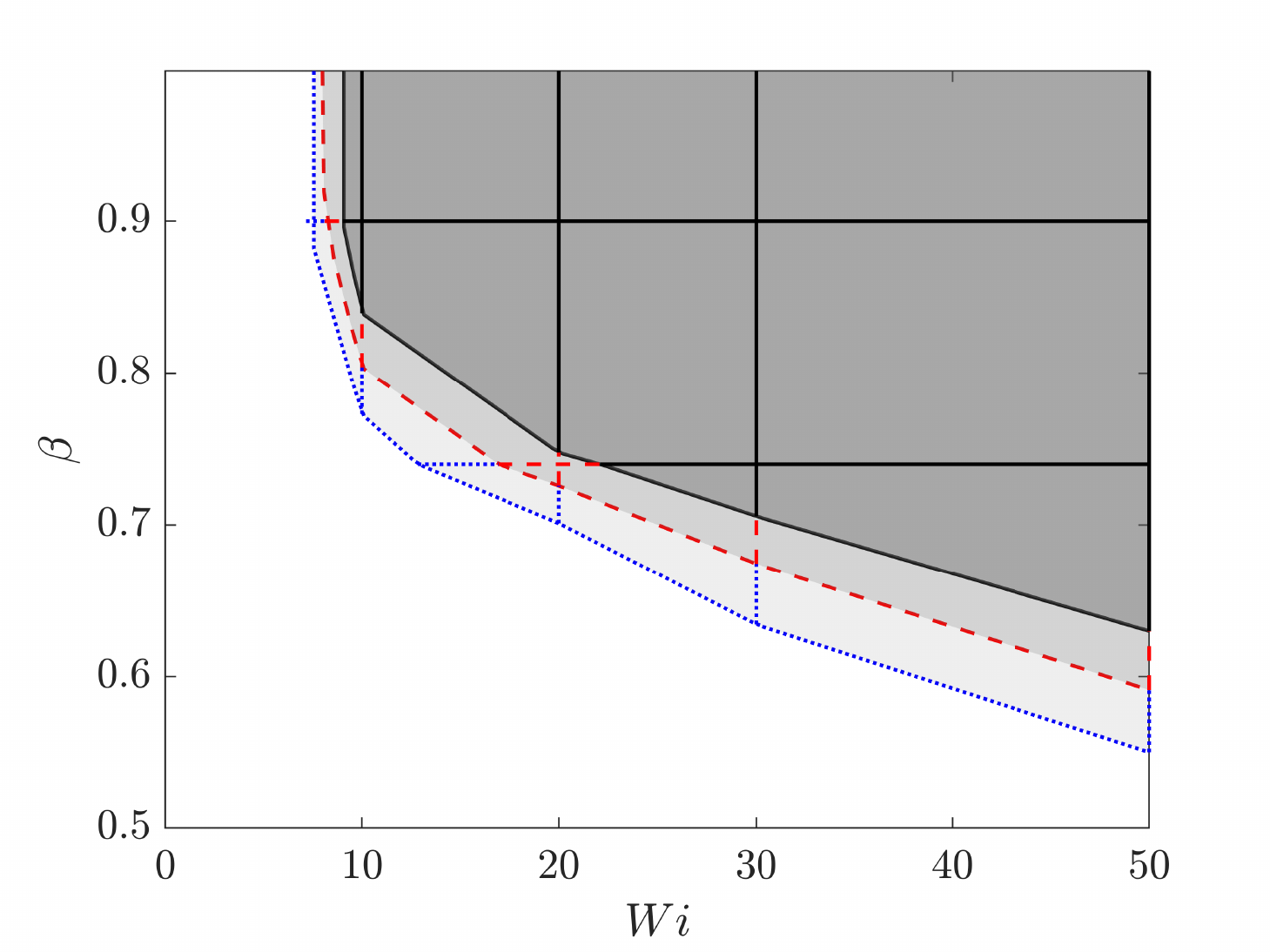}
\centering
\caption{The effect of varying $L_{max}$ on the nonlinearly unstable region at fixed $\lambda = 0.003$. Horizontal and vertical lines correspond to solution branches computed via the continuation routine with $L_{max} = 70$ (black solid lines), $L_{max} = 100$ (red dashed lines), $L_{max} = 500$ (blue dotted lines).}
\label{fig:Wibeta_Ls}
\end{figure}

In addition to the solution branch, Figure~\ref{fig:Au} displays two state plots at $Wi = 20$ and $Wi = 407$, now in terms of $\hat{u}_X/u_{b,X}$ to aid comparison with experimental results.
The former, at $Wi = 20$, still resembles the structural composition of the linearly unstable center eigenmode at higher $Re$ -- chevron shaped streaks remain visible, but are now disconnected at the centerline. By $Wi = 407$ (and similarly for all $Wi>100$ lower branch states), these structures straighten out and form
stream-wise counter-propagating streaks placed symmetrically around the centerline, reminiscent of the structures observed experimentally by \citet{jha2020} and \citet{,Shnapp2021} as coherent structures in ET (see Figure~2 of \citet{jha2020} in particular).
These similarities indicate that the observed stream-wise travelling waves are indeed dynamically connected to the center-mode instability, via a standard subcritical route. 
However, given that our analysis is 2D, we are unable to identify the span-wise travelling waves of \citet{Shnapp2021}, which might serve to explain the discrepancies in threshold amplitude $\mathcal{A}$ on Figure~\ref{fig:Au} -- for instance, there might exist a branch of spanwise travelling solutions with a lower threshold amplitude for $Wi$ large enough. This is under investigation. 

Another recent experimental work observing instabilities in the elastic regime, \citet{Choueiri2021}, was conducted in a pipe - precluding direct comparisons with channel computations performed here -
at even lower viscosity ratios, $\beta \in [0.5,0.6]$.
Without externally perturbing the system, \citet{Choueiri2021} detected fluctuations at $Re \approx 5$ for $\beta = 0.57$, $Wi = 104$ -- however, the flow remains laminar at $Re \approx 3$.
Computationally, the $Re=0$ branch may be continued up to $\beta = 0.57$ (for $\lambda$ sufficiently small) in our channel flow setting (cf.\ Figure~\ref{fig:Wibeta} again), albeit with a slightly larger threshold amplitude than that of Figure~\ref{fig:Au}, which might serve as an explanation for the finite $Re$ required for 'unperturbed' instabilities \citep{Choueiri2021}.
Surprisingly, flow states in \citet{Choueiri2021} still retain their connected, chevron shaped streaks from the center-mode eigenfunction.
In our setting, 'connectedness' of the chevrons is lost shortly after leaving the initial bifurcation (the general shape is still retained for moderate $Wi$, see the middle panel of Figure~\ref{fig:Au}), but the scenario could be quite different in pipes.

\subsection{$L_{max} \in  \{70,100,500\}$}

For completeness, the effect of varying the last outstanding parameter, $L_{max}$, on the region of nonlinear instability is depicted in Figure~\ref{fig:Wibeta_Ls}. 
Contrary to the observations of \citet{buza2021}, which indicated that lowering $L_{max}$ has a destabilizing role in the elastically-dominated regime (cf.\ their Figure~13), here we see that the nonlinearly-unstable region shrinks with decreasing $L_{max}$.
This suppressing effect is in line with past observations of the impact of finite extensibility , e.g.\ see the linear analyses in \citet{Ray2014,Page2015} or even Figure~16 of \citet{buza2021}.
It should be noted that, in our experience, solution branches became difficult to extract at lower $L_{max}$, with convergence issues appearing along upper branches. 
In fact, we can only reliably obtain upper branches for $L_{max}>150$, but lower branches remain accessible due to their lower resolution requirements (see appendix~\ref{B}).

%
%

\section{Discussion}

In this paper, we have used branch continuation to track two-dimensional, finite-amplitude travelling waves in a viscoelastic channel flow, using the FENE-P model.
The travelling waves are borne out of the centre mode instability \citep{Khalid2021a} which is known to be subcritical over large areas of the $Re-Wi$ parameter space \citep{buza2021}. 
Here, we have shown that the TW solution branches extend to significantly lower $Re$ and $Wi$ than the curve of marginal stability. 
For instance, we showed that the saddle node at $Re=1000$ drops as low as $Wi\approx 25$, while the associated linear instability at this $Re$ does not occur until $Wi\approx 80$. 
Most significantly, we demonstrated the persistence of the nonlinear TWs at $Re=0$ for a large range of $Wi$, with the saddle node sitting at $Wi \approx 7.5-10$ in dilute ($\beta=0.9$) solutions for a range of $L_{max}$. 
Across a broad range of the parameter space, including at $Re=0$, the upper branch TWs resemble the arrowhead structures observed in DNS at higher $Re$ \citep{Dubief2020,Page2020}. 

A key feature of the solution branch at $Re=0$ is the low amplitude of the lower branch TW across a \emph{very} large range of $Wi$. 
This suggests that only a very weak disturbance would be required to cause the flow to transition to the higher-amplitude upper branch solution, a scenario which would potentially be indistinguishable from a linear instability in an experiment. 
This observation is consistent with recent experiments showing finite amplitude states in near-inertialess channel flows at very high $Wi$ \citep{jha2020,Shnapp2021}. The small amplitude of the lower branch also poses the question whether the amplitude expansion pursued in \cite{Morozov2005} and \cite{Morozov2019} was attempting to resolve it: e.g. see figure 8(c) of \cite{Morozov2019}. 

While our results demonstrate the existence of finite-amplitude arrowhead TWs over a very large range of the parameter space, a direct connection to either EIT or ET has yet to be established. In inertia-dominated EIT, instantaneous arrowhead-shaped flow structures resembling the stable upper branch states have been observed numerically \citep{Dubief2020}. 
While the region of nonlinear instability found here overlaps with these observations, further work is required to assess if there is a direct connection between the exact coherent states and EIT (e.g. through a sequence of successive bifurcations).  At low-$Re$, our numerical experiments indicate that the upper branch TWs are linearly stable, and we have been unable to trigger ET. Our computations have been restricted to 2D flows, and so this does not rule out the existence of ET in three dimensions, or a direct route from the upper branch TWs to such a state.  We hope to report results on this soon.

\vspace{0.5cm}
\noindent
Acknowledgements: GB gratefully acknowledges the support of the Harding Foundation through a PhD scholarship (https://www.hardingscholars.fund.cam.ac.uk). MB, RK and JP thank EPSRC for support under grant EP/V027247/1.

\vspace{0.5cm}

\noindent
Declaration of Interests. The authors report no conflict of interest.

\vspace{1cm}

%
%
\FloatBarrier
\appendix

%
%
\section{\label{A}Pseudo-arclength continuation}

The branch continuation routine is launched from a point on the neutral curve (curve of marginal linear stability) with the aid of the weakly nonlinear theory, which provides the very first initial condition near the solution branch of interest.
Beyond this point, as parameters are varied in larger increments, supplying sufficiently accurate initial conditions for \eqref{eq:Feq} amounts to predicting the shape of the bifurcation branch,
precisely the objective of pseudo-arclength continuation \citep[see e.g.][]{dijkstra_14}. 
If $\kappa \in \{Wi, Re, \beta\}$ is the parameter allowed to vary in the  continuation (all others are fixed), $\mathcal{F}$ restricts to a map $\mathcal{F}: \mathbb{R}^{Q+1} \to \mathbb{R}^Q$ and \eqref{eq:Feq} translates to
\begin{equation}
    \mathcal{F}(\ba,c,F_X,\kappa) = \mathbf{0}. 
    \label{eq:Feq2}
\end{equation}

Arclength based techniques interpret the branch of solutions as a curve $\bgamma : \mathbb{R} \to \mathbb{R}^{Q+1}$ embedded in configuration space, which is spanned by $(\ba,c,F_X,\kappa)$ in this particular case.
Assume now that a steady state has been computed at point $n$ on the branch $\bgamma$, which we write as $\bgamma(t_n) = (\ba_n,c_n,F_{X,n},\kappa_n)^T$.
A prediction for the solution at the next step, $\bgamma(t_{n+1})$, is given by moving a distance of $s$ tangentially along $\bgamma$, where $s$ denotes the a priori specified step size.
Making use of the fact that a solution branch must satisfy $\mathcal{F} \circ \bgamma \equiv \mathbf{0}$, the tangent at point $n$ is computed according to
$$
\begin{pmatrix}
D \mathcal{F} (\bgamma (t_n)) \\ \dot{\bgamma}(t_{n-1})^T
\end{pmatrix}
 \dot{\bgamma} (t_n) = 
 \begin{pmatrix}
 \mathbf{0} \\ 1
 \end{pmatrix}.
$$
If no tangent is available at the previous step, the last row is replaced by $(0,\ldots,1)$.
The initial prediction for the solution at step $n+1$ is given by
\begin{equation}
    \tilde{\bgamma}^0 (t_{n+1}) := 
    \begin{pmatrix}
\ba_{n+1}^0 \\ c_{n+1}^0 \\ F_{X,n+1}^0 \\ \kappa_{n+1}^0
\end{pmatrix}
= \bgamma(t_n) + s \dot{\bgamma} (t_n),
\label{eq:prediction}
\end{equation}
where the upper indices correspond to the number of completed Newton-Raphson iterates (see below), and the twiddle serves to distinguish the converged branch $\bgamma$ from its approximate counterpart $\tilde{\bgamma}$.
The next step is to employ the Newton-Raphson method using \eqref{eq:prediction} as initial condition to obtain an exact solution of \eqref{eq:Feq2}.
Knowledge of the tangent may be used to aid this procedure, by means of constraining
subsequent iterates to remain on the hyperplane orthogonal to $\dot{\bgamma} (t_n)$.
Incorporating this condition into a standard Newton-Raphson scheme, we obtain the system
$$
    \begin{pmatrix}
        D \mathcal{F}(\tilde{\bgamma}^i (t_{n+1}) ) \\
        \dot{\bgamma}(t_n)^T
    \end{pmatrix}
    \Delta \tilde{\bgamma}^{i+1}(t_{n+1})
    =
    \begin{pmatrix}
        -\mathcal{F}(\tilde{\bgamma}^i (t_{n+1})) \\
        s- \langle \dot{\bgamma}(t_n) , \tilde{\bgamma}^i (t_{n+1}) - \bgamma(t_n) \rangle
    \end{pmatrix},
$$
supplemented with the update rule
$$
\tilde{\bgamma}^{i+1} (t_{n+1}) = \tilde{\bgamma}^i (t_{n+1})+\Delta \tilde{\bgamma}^{i+1}(t_{n+1}).
$$
Once the solutions have converged to a sufficient degree, i.e.,
$$
\frac{|\Delta \tilde{\bgamma}^{i+1}(t_{n+1})|}{|\tilde{\bgamma}^i (t_{n+1})|}< \mathrm{tol},
$$
is reached, we set $\bgamma(t_{n+1}) = \tilde{\bgamma}^{i+1}(t_{n+1})$, and continue further along the branch.
The tolerance $'\mathrm{tol}'$ was set to $10^{-8}$ throughout all examples shown.
The continuation routine consists of repeated applications of the above procedure over $n$, resulting in a discrete representation of the full solution branch as $\{\bgamma (t_n)\}_{n \geq 0}$.

%
%
\section{\label{B} Resolution}
%
%
\begin{figure}
\includegraphics[width=0.48\textwidth]{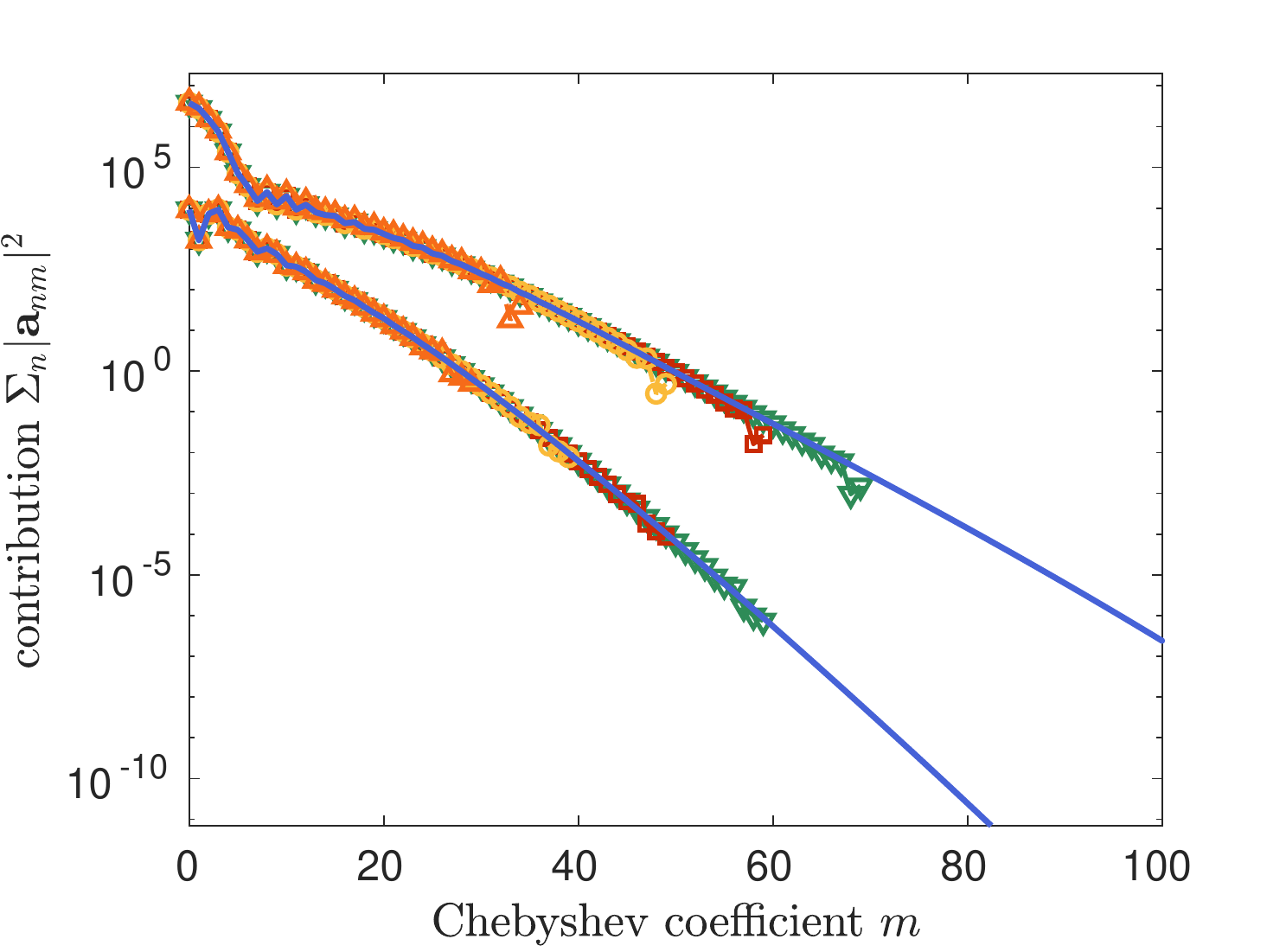}
\includegraphics[width=0.48\textwidth]{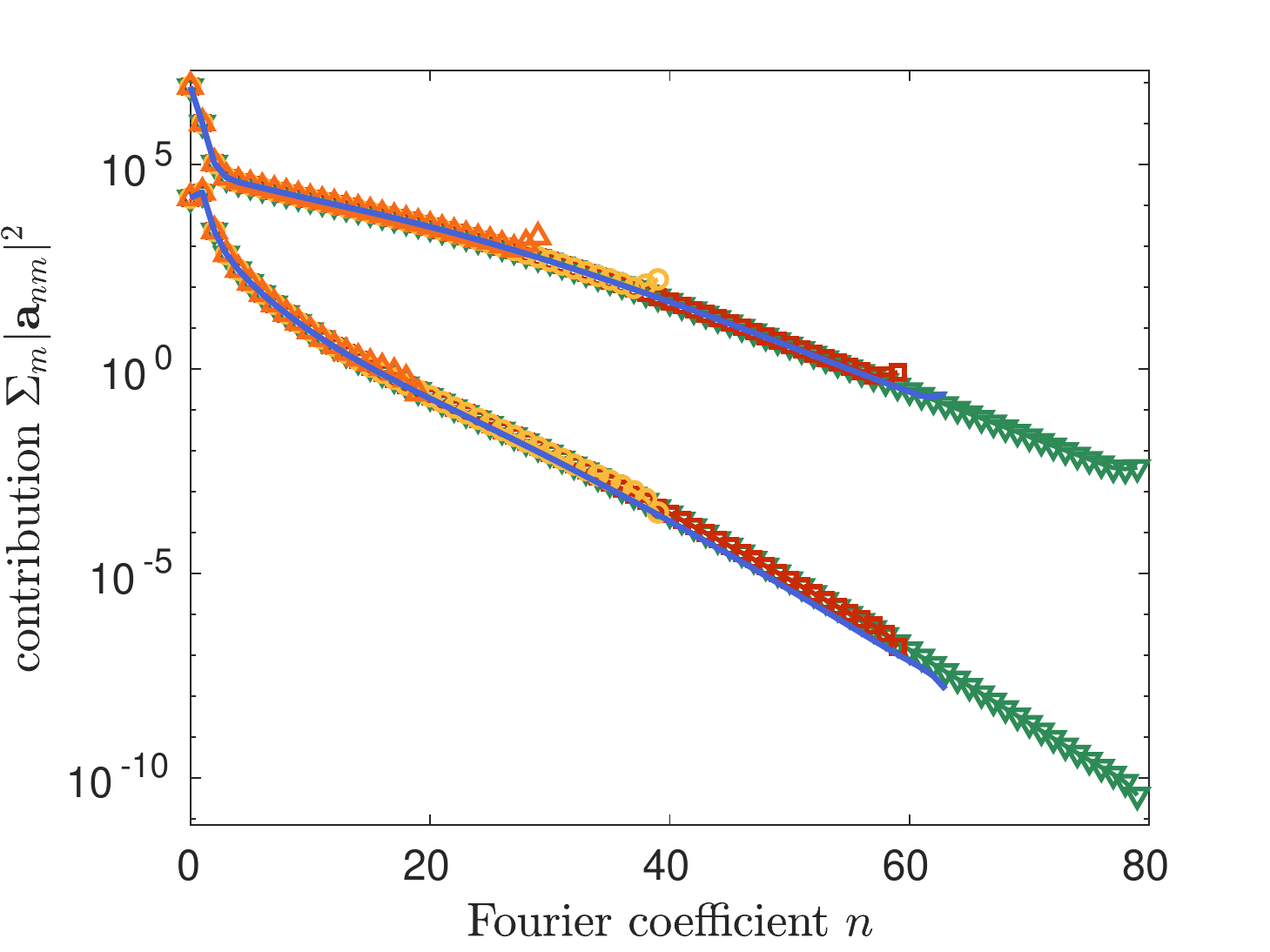}
\centering
\caption{Grid independence checks at $Re = 60$, $Wi = 20$, $L_{max} = 500$, $\beta = 0.9$, $Sc = 250$ for both the upper and lower branch solutions. The upper branch is identified by larger coefficient contributions. Orange, yellow, red and green lines with symbols correspond to solutions obtained through branch continuation, the blue solid lines correspond to the DNS detailed in \S\ref{sec:dns_appendix} using $N_X=128$ complex Fourier modes and $N_y=128$ Chebyshev coefficients. (Left) $N_y$ independence: Overall contribution of each Chebyshev mode for $N_y\in\{30,40,50,60\}$ for the lower branch and $N_y\in\{35,50,60,70\}$ for the upper branch, with $N_X = 60$ fixed in the branch continuation. 
(Right) $N_X$ independence: Overall contribution of each Fourier mode for $N_X\in\{20,40,60,80\}$ for the lower branch and $N_X\in\{30,40,60,80\}$ for the upper branch, with $N_y = 60$ fixed in the branch continuation.}
\label{fig:grid}
\end{figure}

Figure \ref{fig:grid} shows how different truncations in the branch continuation and DNS compare when resolving the upper and lower travelling waves shown in Figure \ref{fig:edge} at $Wi=20$ (more precisely $(Wi,Re,\beta,Sc, L_{max})=(20,60,0.9,250,500)\,$) with the base flow subtracted. The various truncations of branch continuation  show very good agreement with the spectra from the DNS. It is worth remarking that to obtain converged spectra in the DNS it was necessary to reduce the timestep to $\Delta t=1\times 10^{-3}$. Further discrepancies for the higher order modes may be caused by the finite accuracy of the edge tracking algorithm in locating the lower branch.

%
%
%
\section{Remarks on the polymer diffusion term}
\label{sect:poldiff}

Memory limitations associated with increasing $N_X$ and $N_y$ provide an upper bound on how large a value of  $Sc$ can be considered whereas taking $Sc$ too small is known to eradicate small scale dynamics. Thus, $Sc$ was selected between the two of these bounds: just above the smallest number where solution branches become independent of $Sc$, but resolution requirements are still moderate enough for the physical memory to handle. 

In the creeping flow limit ($Re \rightarrow 0$), a distinguished limit with $Sc \rightarrow \infty$ such that $Sc=1/(\varepsilon Re)$ with $\varepsilon$ a constant clearly needs to be taken to retain finite polymer diffusion as follows
$$
    \partial_t \C + \left( \bu \cdot \bnab\right) \C + \T(\C) = \C \cdot \bnab \bu +   \left( \bnab \bu \right)^{T} \cdot \C + \varepsilon \Delta \otimes \C.
$$
The new polymer diffusion coefficient, $\varepsilon$, is selected analogously to $Sc$, whereby independence of solution branches is sought while keeping the grid size manageable.
Numerical observations indicate that there cannot be a universal $\varepsilon$ that is optimal in this sense across all parameter regimes within the $Re=0$ limit.
However, the slight rescaling $\lambda = \varepsilon Wi$ (cf.\ Section~\ref{sect:highelast}) 
seems to help.
%
%
%
\begin{figure}
\includegraphics[width=0.85\textwidth]{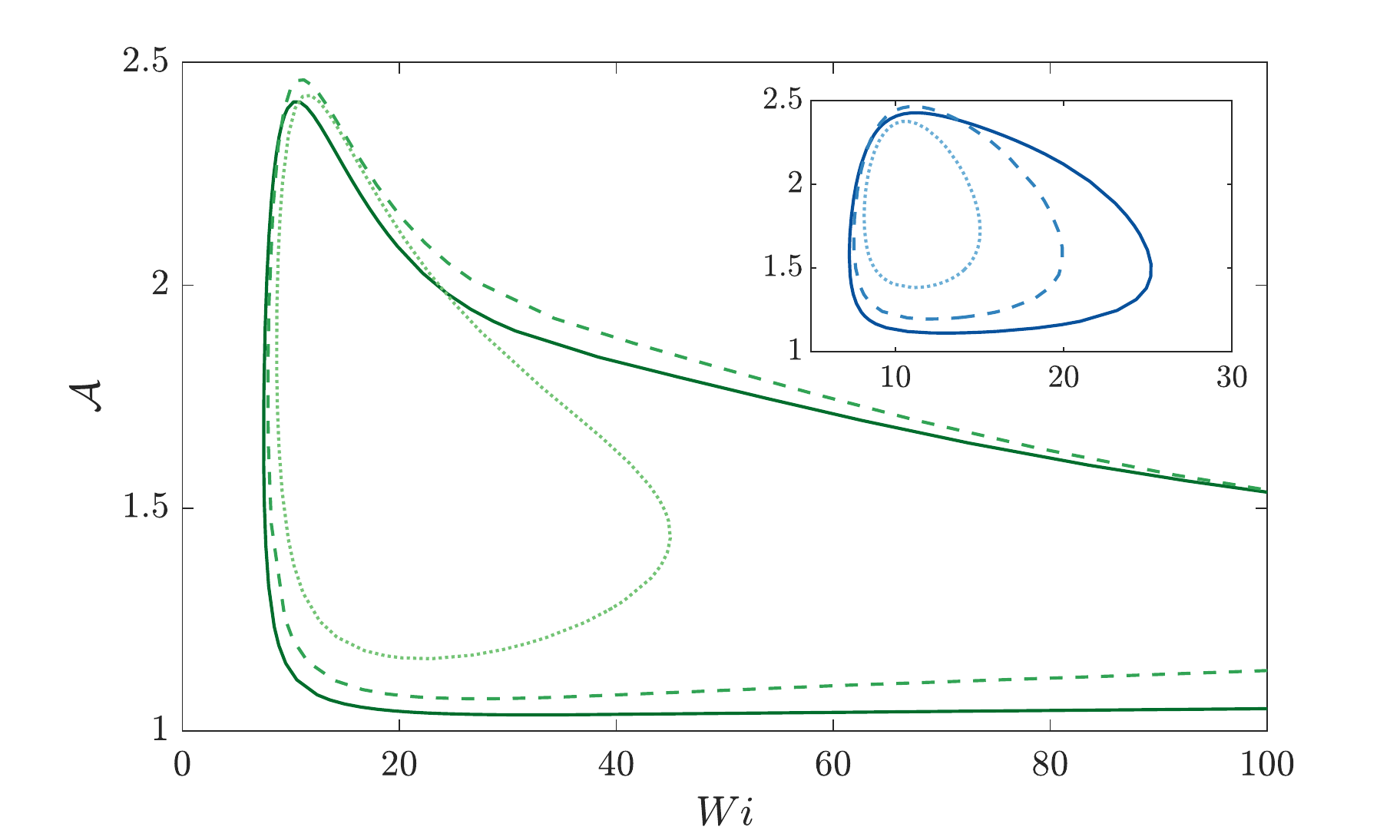}
\centering
\caption{Solution branches at fixed $\lambda = 0.005$ (\sampleline{}), $\lambda = 0.007$ (\sampleline{dashed}), $\lambda = 0.01$ (\sampleline{dotted}). (Inset) Branches with $\varepsilon$ fixed, at $\varepsilon = 5 \cdot 10^{-4}$ (\sampleline{}), $\varepsilon = 7 \cdot 10^{-4}$ (\sampleline{dashed}), $\varepsilon = 10^{-3}$ (\sampleline{dotted}), so that they agree with their $\lambda$-based counterparts at $Wi = 10$.}
\label{fig:eps_lambda}
\end{figure}
Figure~\ref{fig:eps_lambda} displays three pairs of solution branches computed at different values of fixed $\lambda$ (in green) and fixed $\varepsilon$ (in blue).
While the $\lambda$-based branches can be extended up to arbitrarily large $Wi$ (the $\lambda = 0.005$ one was continued up to $Wi > 1000$ - not shown), the $\varepsilon$ based ones form isolas at low Weissenberg numbers $Wi < 30$, resulting in a loss of robustness with respect to $\varepsilon$.
Using $\lambda$, robustness is recovered after a certain threshold is hit ($\lambda \approx 0.007$ for this branch), which served as the primary motivation behind choosing $\lambda$ as the parameter to be fixed throughout the main text.

\bibliographystyle{jfm}


\expandafter\ifx\csname natexlab\endcsname\relax
\def\natexlab#1{#1}\fi
\expandafter\ifx\csname selectlanguage\endcsname\relax
\def\selectlanguage#1{\relax}\fi 

\bibliography{jfm}


\end{document}